\documentclass[11pt, a4paper]{article}
\pdfoutput=1

\usepackage[height=8.85in,width=6.75in]{geometry}

\usepackage{graphicx,rotating}     %usepackage without driver option
\usepackage[bookmarksopen,colorlinks=true,linkcolor=light_blue,
citecolor=light_pink,urlcolor=dark_red,linktoc=all]{hyperref}

\usepackage{amsmath}
\usepackage{mathtools}
\usepackage{amsfonts}
\usepackage{amssymb}
\usepackage{slashed}
\usepackage{braket}
\usepackage{bm}
\usepackage{bbm}
\usepackage{cite}
\usepackage{comment}
\usepackage{xcolor}
\usepackage[utf8]{inputenc}
\usepackage[footnotesize]{caption}
\usepackage{bbold}
\usepackage{xfrac}
\usepackage{physics}

\usepackage{fourier}

\newcommand{\Mpl}{M_{\text{Pl}}}

\makeatletter
\@addtoreset{equation}{section}

\makeatother

\usepackage{multicol}
\definecolor{dark_red}{rgb}{0.7, 0., 0.}
\definecolor{light_pink}{rgb}{1,0.4,0.4}
\definecolor{light_blue}{rgb}{0.284602,0.317763,0.963947}
\definecolor{cred}{RGB}{180,50,40}
\definecolor{darkgreen}{RGB}{0, 100, 0}
\definecolor{desy_blue}{HTML}{009EE2}
\definecolor{desy_orange}{HTML}{FD8800}
\definecolor{forestgreen}{HTML}{228B22}
\definecolor{ochre}{HTML}{CCAA2B}

\newcommand{\eV}{\,\mathrm{eV}}

\newcommand{\MeV}{\,\mathrm{MeV}}
\newcommand{\GeV}{\,\mathrm{GeV}}

\def\ca{C_\text{A}}

\def\df{d_\text{F}}

\def\ta{t_{\rm A}}
\def\tf{t_{\rm F}}

\newcommand{\si}{s_{\rm inj}}
\newcommand{\Br}{{\rm Br}_s}

\newcommand{\nn}{\nonumber\\}

\begin{document}

\hypersetup{pageanchor=false}
\begin{titlepage}

\begin{center}

\hfill KEK-TH-2443\\
\hfill TU-1165\\

\vskip 0.5in

{\Huge \bfseries
Cascades of high-energy SM particles\\
in the primordial thermal plasma\\
}
\vskip .8in

{\Large Kyohei Mukaida$^{a,b}$, Masaki Yamada$^{c,d}$}

\vskip .3in
\begin{tabular}{ll}
$^a$& \!\!\!\!\!\emph{Theory Center, IPNS, KEK, 1-1 Oho, Tsukuba, Ibaraki 305-0801, Japan}\\
$^b$& \!\!\!\!\!\emph{Graduate University for Advanced Studies (Sokendai), }\\[-.3em]
& \!\!\!\!\!\emph{1-1 Oho, Tsukuba, Ibaraki 305-0801, Japan}\\
$^c$& \!\!\!\!\!\emph{FRIS, Tohoku University, Sendai, Miyagi 980-8578, Japan}\\
$^d$& \!\!\!\!\!\emph{Department of Physics, Tohoku University, Sendai, Miyagi 980-8578, Japan}\\
\end{tabular}

\end{center}
\vskip .6in

\begin{abstract}
\noindent
High-energy standard model (SM) particles in the early Universe are generated by the decay of heavy long-lived particles. The subsequent thermalization occurs through the splitting of high-energy primary particles into lower-energy daughters in primordial thermal plasma. The principal example of such processes is reheating after inflation caused by the decay of inflatons into SM particles. Understanding of the thermalization at reheating is extremely important as it reveals the origin of the hot Universe, and could open up new mechanisms for generating dark matter and/or baryon asymmetry. In this paper, we investigate the thermalization of high-energy SM particles in thermal plasma, taking into account the Landau--Pomeranchuk--Migdal effect in the leading-log approximation. The whole SM particle content and all the relevant SM interactions are included for the first time, \textit{i.e.}, the full gauge interactions of SU(3)$_c\times$SU(2)$_L\times$U(1)$_Y$ and the top Yukawa interaction. The distribution function of each SM species is computed both numerically and analytically. We have analytically obtained the distribution function of each SM species after the first few splittings. Furthermore, we demonstrate that, after a sufficient number of splittings, the particle distributions are asymptotic to certain values at low momentum, independent of the high-energy particles injected by inflaton decay. The results are useful to calculate the DM abundance produced during the pre-thermal phase. An example is provided to illustrate a way to calculate the DM abundance from the scattering between the thermal plasma and high-energy particles in the cascade.
\end{abstract}

\end{titlepage}

\tableofcontents
\thispagestyle{empty}
\renewcommand{\thepage}{\arabic{page}}
\renewcommand{\thefootnote}{$\natural$\arabic{footnote}}
\setcounter{footnote}{0}
\newpage
\hypersetup{pageanchor=true}

%%%%%%%%%%%%%
\section{Introduction}
\label{sec:intro}
%%%%%%%%%%%%%

Understanding of the thermal history of the Universe has dramatically changed our perception of cosmology.
Observation of the cosmic microwave background~\cite{Penzias:1965wn,Mather:1993ij} revealed that our Universe was filled with the hot thermal plasma up to $\mathcal{O} (1) \eV$.
Combining this observation with Big Bang nucleosynthesis (BBN), our understanding has been further enhanced to a higher temperature~\cite{Schramm:1997vs}.
In BBN, the abundance of primordial light elements is predicted by solving the Boltzmann equations for nucleons with complicated reaction chains.
The consistency with observed light-element abundances has confirmed the thermal history of the Universe up to a temperature as high as $\mathcal{O}(1)\MeV$~\cite{Kawasaki:1999na,Kawasaki:2000en,Giudice:2000ex,
Hannestad:2004px,Ichikawa:2005vw,deSalas:2015glj,Hasegawa:2019jsa},
providing one of the most stringent constraints on cosmological scenarios beyond the standard model (SM).

To explain the initial condition challenges of the thermal Universe, such as the flatness and horizon issues, an exponentially expanding era, called inflation, must be present in the earlier stage of the Universe~\cite{Guth:1980zm} (see also Refs.~\cite{Starobinsky:1980te,Sato:1980yn}).
When the inflation terminates, its energy should be released into the thermal plasma to form the hot Universe.
This process is called reheating and is realized by the perturbative or nonperturbative decay of inflatons into radiation, including SM particles.
If the decay rate of inflaton is not significantly high,
the last stage of reheating is described by its perturbative decay after a period of inflaton-oscillation domination.%
\footnote{
In the earlier stage of reheating, parametric resonance may take place~\cite{Traschen:1990sw,Kofman:1994rk,Shtanov:1994ce,Kofman:1997yn},  which is known as preheating.
In this paper, we focus on the case with a small inflaton-decay rate, where the preheating is generically shut off by the cosmic expansion and rescatterings.
The final state of reheating is then dominated by the perturbative inflaton decay.
}
The process of thermalization, even in this case, is quite non-trivial, contrary to the naive expectation.
This is because the primary particles injected by the inflaton decay can have much higher energy than the temperature of the ambient plasma, which is as low as $\mathcal{O} (1) \MeV$.\footnote{
  For a higher decay rate, although the reheating is described by the perturbative decay, the temperature of the ambient plasma becomes larger than the inflaton mass.
  In this case, the decay rate is modified by the thermal effect~\cite{Mukaida:2012qn,Mukaida:2012bz}.
}

Moreover, such injection of high-energy SM particles is expected in many models beyond the SM.
If the model involves some heavy long-lived particles, such as dilaton or moduli fields~\cite{Coughlan:1983ci,deCarlos:1993wie,Banks:1993en}, its prolonged decay generates high-energy particles.
Of course, it is not limited to the decay of heavy particles.
The decay/evaporation of extended objects, such as I-balls/oscillons~\cite{Hertzberg:2010yz,Kawasaki:2013awa,Saffin:2016kof,Hong:2017ooe}
and primordial black holes~\cite{Hawking:1974rv,Hawking:1975vcx,Page:1976df,Das:2021wei}, also lead to such primary generation of high-energy particles.
To have the thermal Universe of $\mathcal{O} (1) \MeV$ confirmed by the success of BBN, one has to guarantee that the primary high-energy particles are thermalized until then.
Therefore, understanding of the thermalization process after the injection of high-energy SM particles is indispensable not only to cosmology but to particle physics.

High-energy SM particles are continuously injected into low-temperature plasma before the completion of, for instance, inflaton decay.
In order for the high-energy particles to become thermalized, they must lose their energy while increasing in number via their splittings into lower momentum modes.
The splittings of parent particles with much higher energy than the temperature of the ambient plasma occur almost collinearly, and hence, the rate is strongly suppressed by the interference between the parent and daughter particles.
This effect is known as the Landau--Pomeranchuk--Migdal (LPM) effect~\cite{Landau:1953um, Migdal:1956tc, Gyulassy:1993hr,
 Arnold:2001ba, Arnold:2001ms, Arnold:2002ja, Besak:2010fb}.
The resulting thermalization proceeds via the bottom-up process, where lower-momentum modes are thermalized earlier,
and the injected high-energy particles are thermalized only after the splittings.
Thermalization was investigated in detail
in Refs.~\cite{Arnold:2002zm,Jeon:2003gi, Arnold:2008zu, Kurkela:2011ti,AbraaoYork:2014hbk,Kurkela:2014tea,Kurkela:2014tla,Kurkela:2018oqw,Kurkela:2018xxd,Du:2020zqg,Du:2020dvp,Fu:2021jhl} in the context of ultra-relativistic heavy-ion collisions, which were applied to the cosmological context in
Refs.~\cite{Harigaya:2013vwa,Harigaya:2014waa,Mukaida:2015ria,Drees:2021lbm,Passaglia:2021upk,Drees:2022vvn} (see also Refs.~\cite{Davidson:2000er,Allahverdi:2002pu,Jaikumar:2002iq} for earlier works).

To calculate the amount of non-thermally produced dark matter (DM) during thermalization and reheating, understanding of detailed thermalization history is crucial.
In particular, during splitting, the number of high-energy particles exponentially increases (though their energy becomes smaller).
Subsequently, weakly interacting dark matter can be efficiently produced from the collisions of high-energy particles before they are completely thermalized~\cite{Harigaya:2014waa}.%
\footnote{
Non-thermal production of DM during reheating and thermalization was discussed in Refs.~\cite{Garcia:2017tuj, Dudas:2017kfz, Drees:2017iod, Allahverdi:2018aux, Kaneta:2019zgw,Bernal:2019mhf,Allahverdi:2019jsc}.
However, the following effects were not taken into account: finiteness of the thermalization timescale and DM production from the thermal cascade of high-energy particles.
See also Refs.~\cite{Chung:1998rq, Giudice:2000ex, Allahverdi:2002nb, Allahverdi:2002pu, Kane:2009if, Hooper:2011aj,Kurata:2012nf, Fan:2013faa, Kane:2015jia, Co:2015pka, Dhuria:2015xua} for earlier works on non-thermal production of DM with and without the instantaneous thermalization approximation.
See also Refs.~\cite{Kurata:2012nf,Mambrini:2022uol} for indirect non-thermal production of DM from the inflaton decay in vacuum.
}
The following related works included a non-renormalizable coupling for DM production~\cite{Garcia:2018wtq,Harigaya:2019tzu}%
\footnote{
See Ref.~\cite{Garcia:2020eof,Garcia:2020wiy} for the case with a non-quadratic inflaton potential.
}
non-thermal leptogenesis during thermalization ~\cite{Hamada:2015xva, Hamada:2018epb, Asaka:2019ocw},
and sphalerons after the electroweak crossover~\cite{Asaka:2003vt,Jaeckel:2022osh}.
In Ref.~\cite{Drees:2022vvn}, the Boltzmann equations were numerically solved for fermions and gauge fields in the SM without the Higgs field, with the results agreeing with qualitative discussion~\cite{Harigaya:2014waa} and numerical results for a pure gluon theory~\cite{Drees:2021lbm} up to some numerical factors.

In this paper, we extend the analysis and calculation of Refs.~\cite{Harigaya:2014waa,Drees:2021lbm,Drees:2022vvn} by providing and solving complete Boltzmann equations for the SM particles in the leading-log approximation for the thermalization of high-energy SM particles. We include the quarks, leptons, Higgs, and gauge bosons of SU(3)$_c\times$SU(2)$_L\times$U(1)$_Y$ gauge theory. The top Yukawa interaction is considered, while the other Yukawa interactions are negligible.
The complete Boltzmann equations are numerically solved in the stationary regime
where the source term of high-energy particles is balanced by the dissipation into the thermal plasma.
This corresponds to the thermalization at the last stage of reheating, where the LPM suppressed splitting rate is much larger than the Hubble expansion rate~\cite{Mukaida:2015ria}.
The relative values of distributions of the SM fields at a given momentum are observed to be asymptotic to certain values at low momenta, independent of the initially injected high-energy particles.
Therefore, all SM particles are produced during the splitting process and their distributions reach the scaling solution in the limit of a large number of splitting processes.
The energy scale at which the scaling solution is approximately realized is determined, which depends on the initially injected particle species.
The results can be used to consider the non-thermal production of DM during thermalization.

This paper is organized as follows:
In Sec.~\ref{sec:reheating}, first, the system in which our calculations are applied is specified. The properties and Boltzmann equations are summarized for the SM particles that govern the thermalization after inflation.
In Sec.~\ref{sec:analytic}, we provide some analytic results for the asymptotic behavior of the distribution functions for the SM particles at a small and large momentum.
The former provides a scaling solution of the Boltzmann equation that is useful for calculating the DM production process during thermalization.
Subsequently, the scaling behavior at a large momentum provides an appropriate boundary condition for numerical calculation of the Boltzmann equation.
Equipped with this boundary condition, the Boltzmann equation is numerically solved in Sec.~\ref{sec:numerical}.
The numerical result confirms the analytic asymptotic behavior of the distribution functions.
In Sec.~\ref{sec:application}, the results are applied to the non-thermal production of DM.
A toy model is considered to illustrate the calculation of the DM abundance from scattering between the thermal plasma and a cascading high-energy particle.
Section~\ref{sec:conclusion} presents the discussion and conclusion.

%%%%%%%%%%%%%
\section{Setup for thermalization process of SM particles}
\label{sec:reheating}
%%%%%%%%%%%%%

\subsection{Source term for primary particles}

We are interested in the thermalization process of high-energy particles injected into a low-temperature ambient plasma.
This is realized, \textit{e.g.}, by the decay of inflaton into SM particles during the reheating epoch.
Our calculation and formalism can be applied to a more general setup, which we specify below.

The system is as follows:
We begin with a thermal plasma with a temperature $T$. High-energy SM particles with energy $p_0$ ($\gg T$) are injected into the thermal plasma via, \textit{e.g.}, the decay of a heavy particle.
We refer to these high-energy particles as primary particles.
These primary particles are expected to lose their energy via the splitting process into lower-energy daughter particles, as explained later in this paper.
This process can be described by the Boltzmann equations if we appropriately take into account the splitting processes as we discuss in the next Sec.~\ref{sec:splitting}.
\begin{equation}
 \qty( \frac{\partial }{\partial t} - H p \frac{\partial }{\partial p} ) f_s(p,t)= \text{(Source \ term)} + \text{(Splitting \ terms)},
 \label{eq:Boltzmann0}
\end{equation}
where $H$ is the Hubble parameter and $f_s$ represents the distribution function of particle species $s$. The source term is present for the primary particles and is given by a delta-function at $p = p_0$.
If we consider a case in which the primary particles originate from the two-body decay of a heavy particle with number density $n_I (t)$, mass $m_I$, and decay rate $\Gamma_I$, the source term is expressed as
\begin{align}
 &\text{(Source \ term)} = \Br \frac{\dd \Gamma_I}{\dd p} \frac{2\pi^2}{p^2} n_I(t),
 \\
  &\frac{\dd \Gamma_I}{\dd p}= 2 \Gamma_I \delta \qty( p - p_0 ), \qquad
 p_0 = m_I / 2.
  \label{eq:source}
\end{align}
Here, $\Br$ is the branching ratio into a particle species $s$, which is defined shortly [see Eq.~\eqref{eq:Br}].%
\footnote{
\label{footnote1}
Strictly speaking,
the source term is a distribution with a finite width broadening in the domain of $p \le p_0$
because of, \textit{e.g.}, the redshift via the expansion of the Universe.
This particularly implies that
the integral of the delta function over $p$ from $p \ll p_0$ to $p_0$ gives a factor of $1$ rather than $1/2$.
In other words, the delta-function gives a source only for $p < p_0$ (rather than both $p< p_0$ and $p > p_0$), so that it gives a factor of $1$ after the integral over a line segment of $[0,p_0]$.
}
If the heavy particle denoted by the subscript $I$ behaves as a pressureless matter, we have
\begin{equation}
  n_I(t) = n_I (t_0) \qty[ \frac{a(t_0)}{a(t)} ]^3 e^{-\Gamma_I t} ,
  \label{eq:nI}
\end{equation}
where $a(t)$ is the scale factor, and $t_0$ is a reference time.
The splitting terms will be specified in the next Sec.~\ref{sec:splitting}.

In this paper, we consider a
regime in which the thermalization rate is significantly faster than the Hubble expansion rate,
and hence the temperature of the ambient plasma can be approximated to be constant.
Then, we can neglect the second term on the left-hand side of \eqref{eq:Boltzmann0}.
We also consider the case in which $n_I(t)$ does not change over the thermalization timescale.
For notational convenience, we define
\begin{align}
 &\tilde{\Gamma} \equiv 2 \Gamma_I \frac{2\pi^2 n_I}{p_0^2} \frac{1}{p_0^{1/2} T^{3/2}},
\end{align}
so that
the source term of the Boltzmann equation can be expressed as
\begin{align}
\text{(Source \ term)} = p_0^{1/2} T^{3/2} \Br \tilde{\Gamma} \delta( p -p_0).
\end{align}
The factor of $p_0^{1/2} T^{3/2}$ is included for later convenience.

Since the thermalization timescale is faster than other timescales and primary particles are continuously injected,
the particle distributions are expected to reach a stationary solution.
This greatly simplifies our analysis, as we only need to calculate a stationary solution to the Boltzmann equations.
One may obtain such stationary solution by requiring that the collision terms of the Boltzmann equations vanish for a given source term.

There are several interesting applications for our calculations.
One simple example is the decay of a heavy particle into SM particles in the early Universe.
A particularly important one is reheating via the perturbative decay of inflaton, which is expected at the last stage of reheating
for many inflaton models with a small decay rate such as a Planck-suppressed decay.
Such a small coupling is theoretically well motivated because inflaton must have a flat potential, which is spoiled by loop corrections if the inflaton has sizable interactions.
In this case, the above assumptions are justified when one considers the thermalization of inflaton-decay products at the final stage of the reheating process.%
\footnote{
Strictly speaking, the thermalization process (more specifically, the LPM splitting process) occurs faster than the expansion rate of the Universe after the Universe reaches its maximal temperature, as discussed in Ref.~\cite{Mukaida:2015ria}. One can apply our calculation to this regime.
}
Furthermore, one can approximate $n_I(t)$ to be a constant value because inflaton decay is significantly slower than the thermalization timescale.
Let us illustrate a typical time-evolution after inflation for the sake of completeness.
After inflation, the inflaton oscillation dominates the Universe, and the inflaton perturbatively decays into the SM particles to reheat the Universe.
When the Hubble parameter $H$ becomes comparable to the decay rate $\Gamma_I$, reheating is completed, and the Universe is dominated by radiation.
The important point here is that the inflaton continuously decays into SM particles, \textit{even before reheating is completed} (\textit{i.e.}, during the inflaton-oscillation dominated era).
This is the source of primary particles during the reheating epoch.

Throughout this paper, we assume that the thermalization proceeds in a homogeneous ambient plasma with constant temperature $T$ and can be treated as if  thermalization of each high-energy particle is an isolated event.
This is justified because the number density of decaying heavy particle is small enough for the case of interest
and the backreaction of cascading process to ambient plasma is negligible.
This is confirmed by the following order-of-magnitude estimation.
By each thermalization process,
a small region is heated via the splitting process which we will explain shortly.
Those heated regions
dissipate into the ambient plasma with the diffusion length of order $(t \, t_{\rm el})^{1/2} \sim t^{1/2} / (\alpha T^{1/2})$ within a time scale of $t$, where $t_{\rm el} \sim 1/(\alpha^2 T)$ is the time scale of elastic scatterings.
In order to see the effect of the diffused region of a single jet to the subsequent jets,
the typical time scale of interest should be self-consistently determined such that another jet is produced within the volume of $d_{\rm dif}^3$.
Here, the probability that a heavy particle decays within the volume of $d_{\rm dif}^3$ is given by $d_{\rm dif}^3 n_I \Gamma_I t$.
Setting this equal to unity and solve it in terms of $t$,
we obtain
$t \gtrsim \left( \alpha^3 p_0 \Mpl \right)^{2/5} / T^{9/5}$,
where we used $n_I \lesssim T^4 / p_0$ and $\Gamma_I \lesssim H \sim T^2 / \Mpl^2$.
The volume of dissipated region from each high-energy particle is about $d_{\rm dis}^3$ and the energy of injection into this region is about $p_0$.
Comparing this with the energy of thermal plasma in that region, we can estimate how much the thermalization process overheats this region such as
\begin{align}
 \frac{p_0}{d_{\rm dif}^3} \frac{1}{T^4}
 \lesssim  \left( \frac{\alpha^6 p_0^2 T}{M_{\rm pl}^{3} } \right)^{1/5}.
\end{align}
This is much smaller than unity, so that we can neglect the backreaction to the ambient plasma during the thermalization
and approximate $T$ to be constant in space.

\subsection{Splitting process}
\label{sec:splitting}

\begin{figure}[t]
	\centering
 	\includegraphics[width=0.6\linewidth]{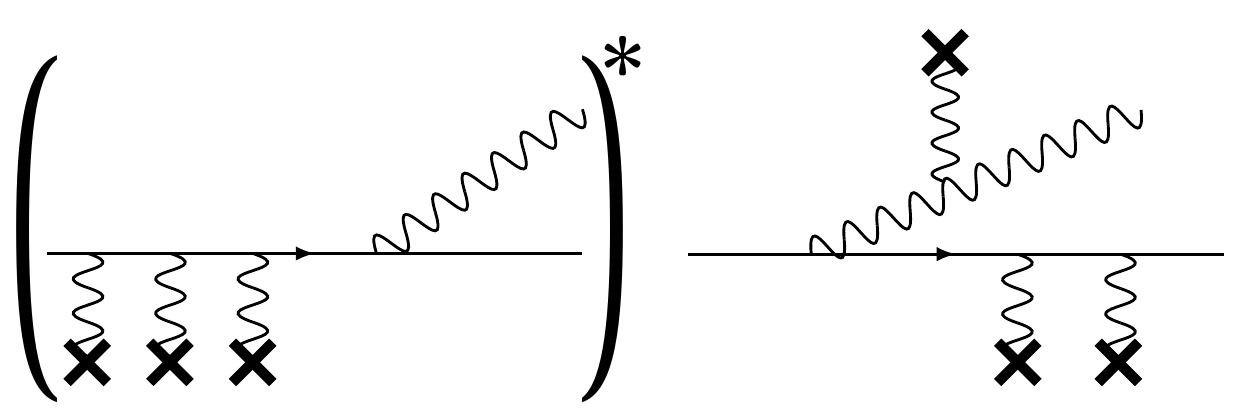}
	\caption{Example of interference between diagrams where the emission occurs before and after multiple scatterings with thermal plasma.
  Here the cross denotes the thermal plasma.
  Such interference leads to the LPM effect.
}
	\label{fig:LPM}
\end{figure}

Our entire analysis is based on Boltzmann equations, which are valid under some conditions.
One can show that the (quasi-)particle excitations with momenta larger than the screening scale, $p^2 \gg m_D^2 \sim \alpha \int_{\bm p} f(p) / |p|$, are described by Boltzmann equations
if the following conditions are met:
(i) the occupancy is perturbative, $f(p) \ll 1 / \alpha$,
(ii) the size of quasi-particles is smaller than the mean free path,
and (iii) the duration of each interaction is shorter than the mean free time.
Here the fine structure constant of relevant interactions is denoted as $\alpha$ collectively.
The last condition implies the quantum decoherence of individual scatterings, which requires treating each interaction as the collision terms in the Boltzmann equations.

We are interested in the energy loss of high-energy particles injected into low-temperature ambient plasma, which is dominated by their nearly collinear splittings.
For such collinear emissions, the last condition (iii) should be discussed carefully because the emitted daughter stays close to the parent, thereby interfering with subsequent scatterings. This is known as the LPM effect~\cite{Landau:1953um, Migdal:1956tc, Gyulassy:1993hr,
 Arnold:2001ba, Arnold:2001ms, Arnold:2002ja, Besak:2010fb}. See Fig.~\ref{fig:LPM} as an illustration.
Hence, coherence is kept until the overlap between the parent and daughter is lost.

Let us estimate the decoherence time, $t_\text{form}$, before which the destructive interference suppresses the emissions following Ref.~\cite{Kurkela:2011ti}.
Suppose that a parent particle with a momentum $p$ emits a daughter particle with a momentum $k$ almost collinearly which is charged under the SM gauge group.
In this case, the decoherence time is dominated by the daughter particle.
Since the transverse size of the wave is $1/k_\perp$ with $k_\perp$ being its transverse momentum, the overlap is resolved for $t \gtrsim k / k_\perp^2$.
A high-energy charged particle acquires transverse momentum through frequent elastic scatterings with particles in the medium mediated by the $t$-channel gauge boson exchange.
In our case of interest, such elastic scatterings can be regarded as random processes because we expect $t_\text{form} \gg t_\text{el}$.
The square transverse momentum then obeys the diffusion equation of
\begin{equation}
    k_\perp^2 \sim \hat q_\text{el} t,
    \qquad
    \hat q_\text{el} \sim \int_{\bm{q}_\perp} \frac{\alpha^2 (q_\perp)}{q_\perp^2 + m_D^2} \int_{\bm{p}'} f(\bm{p}'),
    \qquad
    m_D^2 \sim \alpha \int_{\bm{p}'} \frac{f (\bm{p}')}{p'},
    \label{eq:transverse_diff}
\end{equation}
with $\alpha$ being the fine structure constant of the mediated gauge boson.
This leads to the following condition for the decoherence:
\begin{equation}
  t \gtrsim \sqrt{k / \hat q_\text{el}} \equiv t_\text{form}.
\end{equation}
If the daughter particle is neutral under a gauge group of the system in consideration, the decoherence time is dominated by the transverse diffusion of the parent particle, \textit{i.e.}, $p_\perp^2 \sim \hat q_\text{el} t$.
In this case, we instead have
\begin{equation}
  t \gtrsim \frac{1}{k_\perp} \frac{p}{p_\perp} \simeq \frac{1}{k} \frac{p^2}{p_\perp^2}
  ~~\longrightarrow~~
  t \gtrsim \sqrt{\frac{p^2}{k \hat q_\text{el}}} \equiv t_\text{form}.
\end{equation}
An important example of this case is the emission of U(1)$_Y$ gauge boson.

The above consideration shows that the splitting can only happen for $t > t_\text{form}$.
Once the condition $t > t_\text{form}$ is fulfilled, the subsequent splitting can be treated incoherently, which allows us to use the Boltzmann equations.
Taking into account the coupling to the daughter particle $\alpha_d$, the LPM-suppressed splitting rate is estimated as
\begin{equation}
  \Gamma_\text{LPM} \sim \frac{\alpha_d}{t_\text{form}}
  = \alpha_d \,
  \begin{cases}
    \sqrt{\frac{\hat q_\text{el}}{k}} &\text{charged daughter}, \\
    \sqrt{\frac{k \hat q_\text{el}}{p^2}} & \text{neutral daughter}.
  \end{cases}
\end{equation}
The splitting rate is suppressed in proportion to $1/ \sqrt{k}$ or $\sqrt{k/p^2}$, reproducing the characteristic suppression factor of the LPM effect.
The corresponding splitting function that appears in the Boltzmann equations is given by
\begin{equation}
  \gamma \sim k \times \Gamma_\text{LPM} \sim \alpha_d \left. k_\perp^2 \right|_{t \sim t_\text{form}}, \qquad
  \left. k_\perp^2 \right|_{t \sim t_\text{form}} \sim
    \begin{cases}
      \hat q_\text{el} t_\text{form} = \sqrt{k \hat q_\text{el}} &\text{charged daughter},\\
      \frac{k^2}{p^2} \hat q_\text{el} t_\text{form} = \frac{k}{p} \sqrt{k \hat q_\text{el}} & \text{neutral daughter}.
    \end{cases}
  \label{eq:split_func_rough}
\end{equation}
As we see shortly, this simple argument correctly reproduces the qualitative behavior of the splitting functions.

The exact splitting function with the LPM effect is obtained by resumming the corresponding diagrams.
The resummation can be performed by solving the recursion equation self-consistently, which is derived from first principles in thermal field theory~\cite{Arnold:2001ba, Arnold:2001ms, Arnold:2002ja, Besak:2010fb}.
Instead of providing a direct derivation here, we just quote the result known in the literature and explain the physical meaning of each term.
Suppose that the parent particle of species $s$ with momentum $p$ emits daughters of species $s'$ and $s''$ with momenta of $k$ and $p - k$ almost collinearly.
The splitting functions can be written collectively as
\begin{equation}
  \gamma_{s \leftrightarrow s' s''} \qty( p; xp, (1 - x)p ) = \frac{1}{2} \frac{\alpha_{ss's''}}{\qty( 2 \pi )^4 \sqrt{2}} \times
  \frac{P_{s \leftrightarrow s' s''}^\text{(vac)} \qty(x)}{x \qty( 1 - x)} \times \mu^2_\perp \qty( 1, x, 1-x; s, s', s''),
\end{equation}
with $x = k / p$.
The first term $\alpha_{ss's''}$ collectively represents the coupling of a three-point vertex of $s s' s''$, including the possible group factors.
For instance, the triple gauge boson vertex of SU$(N)$ gives $\alpha_{g g g} = \qty(N^2 - 1) N \alpha$ with $\alpha$ being the fine structure constant of SU$(N)$.
The second term in the numerator is the well-known splitting function in the DGLAP equations.
Hence, the first two terms correspond to the splitting processes in a vacuum.

The last term comes from the diffused transverse momentum at the decoherence time, which is a more general expression of $k_\perp^2 |_{t\sim t_\text{form}}$ in Eq.~\eqref{eq:split_func_rough}.
The explicit form of $\mu_\perp$ for $p \gg T / (\alpha^2 \ln 1/ \alpha^2)$ in the leading-log approximation is given by~\cite{Arnold:2008zu}%
\footnote{
For a relatively small momentum,
$T \ll p_0 \ll T / (\alpha^2 \log 1/\alpha^{1/2})$, one may
replace
\begin {equation}
  \frac{ \alpha_a \qty(m_{D,a}) - \alpha_a \qty(Q_{\perp, a}) }{ - b_a / \qty( 64 \pi^3 )}\, \mathcal{N}_a
  \quad \to \quad
  4\pi \alpha_a m_{D,a}^2 T \ln(Q_{\perp, a}^2/m_{D,a}^2)
\label {eq:logsub}
\end {equation}
in \eqref{eq:mu_perp}.
However, the difference is of the order of $\mathcal{O}(10)\%$ at most, even for such a small momentum; therefore, we use \eqref{eq:mu_perp} throughout our numerical calculations.
}%
\footnote{
Because we are interested in the case with a large hierarchy between $p_0$ and $T$, logarithmic corrections of order $\alpha_a \log (Q_{\perp,a}/T)$ to the leading-log approximation could be important.
This kind of corrections is known as doubly logarithmically enhanced order-$\alpha$ corrections~\cite{Iancu:2015vea}.
We thank an anonymous referee for pointing this out and leave the effect of those corrections for a future work.
On the contrary, $\sqrt{\alpha}$-corrections can be neglected because it is smaller by of order  $\sqrt{\alpha}/\log (Q_{\perp,a}/T)$ than the leading-log term~\cite{Caron-Huot:2008zna}.
}
\begin{align}
  \mu^4_\perp \qty( x_1, x_2, x_3; s_1, s_2, s_3) =
      \frac{2}{\pi} \,
      x_1 x_2 x_3 \,
      p\,
    \sum_{a}  \frac{ \alpha_a \qty(m_{D,a}) - \alpha_a \qty(Q_{\perp, a}) }{ - b_a / \qty( 64 \pi^3 )}\, \mathcal{N}_a
    \sum_{\sigma \in A_3} \tfrac12 \qty[ C_{R_{s_{\sigma(2)}}}^{(a)}+C_{R_{s_{\sigma(3)}}}^{(a)}-C_{R_{s_{\sigma(1)}}}^{(a)} ]  x_{\sigma(1)}^2,
    \label{eq:mu_perp}
\end{align}
where
\begin{equation}
  \mathcal{N}_a
  \equiv
 \sum_s \frac{\nu_s}{d_{{\rm R}_s}^{(a)}} t_{{\rm R}_s}^{(a)}
 \int \frac{\dd^3\ell}{(2\pi)^3} \>  f_s(\ell),
 \label{eq:N_a}
\end{equation}
and
\begin{equation}
  \qty( \frac{Q_{\perp,a}}{m_{D,a}} )^2
  \sim \left(\frac{p}{T}\right)^{1/2} \ln^{1/2}\left(\frac{p}{T}\right) \,,
  \qquad
  m_{D,a}^2 =
    8\pi \alpha_a (T)
    \sum_s \frac{\nu_s}{d_{\mathrm{R}_s}^{(a)}} t_{\mathrm{R}_s}^{(a)} \int\frac{\dd^3p}{(2\pi)^3} \frac{f_s(p)}{p}
  \,.
  \label{eq:mD}
\end{equation}
Here, we consider a gauge group of $G = G_1 \times G_2 \times \cdots \times G_N$ with $G_a$ being a Lie group and the summation over $a$ runs through all the Lie groups involved, \textit{i.e.}, $a = 1 ,2,\cdots,N$.
We use $s$ to represent particle species, its number of degrees of freedom is denoted by $\nu_s$, and the corresponding representation under $G$ is $\mathrm{R}_s$.
For non-Abelian $G_a$, the dimension of a representation $\mathrm{R}$ is $d_\text{R}^{(a)}$, its generators are normalized by $\mathrm{Tr} [T_{{\rm R}}^{i} T_{{\rm R}}^j] = t_{{\rm R}}^{(a)} \delta_{ij}$, and its quadratic Casimir is denoted as $C^{(a)}_{{\rm R}} \cdot \mathbb{1} \equiv T_{{\rm R}}^{i} T_{{\rm R}}^i$,
where $i$ and $j$ are indices for generators.
For notational brevity, we use the same characters $d_\mathrm{R}^{(a)}, t_\mathrm{R}^{(a)}, C^{(a)}_{{\rm R}}$ for Abelian $G_a$, which are defined by $d_\mathrm{R}^{(a)} = 1$ and $t_{{\rm R}}^{(a)} = C^{(a)}_{{\rm R}} = q_{a}^2$ with $q_a$ being the charge under $G_a$.
The fine-structure constant of each $G_a$ is denoted as $\alpha_a$ and its beta function is represented by $b_a$.
Note here that all the gauge couplings are assumed to be comparable $\alpha \sim \alpha_a$ and hence the resummation is performed for all the gauge fields on an equal footing as done for instance in \cite{Anisimov:2010gy,Besak:2012qm,Bodeker:2019ajh}.
The summation is taken over the alternating group of degree $3$ denoted as $A_3$, \textit{i.e.}, $(\sigma(1), \sigma(2), \sigma(3)) = (1,2,3)$, $(2,3,1)$, or $(3,1,2)$.
As an example, let us again consider the case of SU$(N)$ gauge fields.
One finds $\mu_\perp^2 (1, x, (1-x); g g g) \sim \sqrt{xp \hat q_\text{el}}$ for $x \ll 1$.
On the other hand, the emission of a U$(1)$ gauge field from $\psi$ gives
$\mu_\perp^2 (1, x, (1-x); \psi g \psi) \sim x \sqrt{x p \hat q_\text{el}}$.
They coincide with $k_\perp^2 |_{t \sim t_\text{form}}$ in Eq.~\eqref{eq:split_func_rough}.

\subsection{SM particles}

Now we shall consider the SM, where $G = G_1 \times G_2 \times G_3$ with $G_1 =$U(1)$_Y$, $G_2 =$SU(2)$_L$, and $G_3 = $SU(3)$_c$.
We consider the thermalization process of hard primaries via SM interactions
and thus assume that hard primaries are one of the (or some combinations of) SM particles.
We denote the primary particle as $\si$.

We first summarize the properties of SM particles for the reader’s convenience.
A list of SM particles and their charge assignments in SU(3)$_c\times$SU(2)$_L\times$U(1)$_Y$ are shown below.
\begin{center}
\begin{tabular}{c|cccccc|ccc}
	& $e_f$ & $L_f$ & $u_f$ & $d_f$ & $Q_f$ & $\phi$ & $g$ & $W$ & $B$  \\[.2em]
	\hline
	SU(3)$_c$ &   &  & F & F & F &
	& A &  &
	\\
	SU(2)$_L$ &  & F &   &   & F & F
	& & A &
	\\
	U(1)$_Y$ & -1 & -1/2 & 2/3 & -1/3 & 1/6 & 1/2
	&  &  &
	\\
	\hline
	$\nu_s$ & 2 & 4 & 6 & 6 & 12 & 4 & 16 & 6 & 2
\end{tabular}
\label{tab:toolkit}
\end{center}
where $f= 1,2,3$ represents the family
and $\nu_s$ represents the number of degrees of freedom of the particle species $s$.
We neglect asymmetry in the system, that is, the number density of an anti-particle is equal to that of a particle, so that its degrees of freedom are included in $\nu_s$.
We use $s$ to represent particle species, such as
\begin{equation}
 s = \qty( e_f, L_f, u_{f'}, u_3, d_f, Q_{f'}, Q_3, \phi, g, W, B ),
\end{equation}
where $f=1,2,3$, and $f'=1,2$.
In our notation, when we take a summation over an index for species $s$, we implicitly include that of $f$.
We treat the third family of quarks separately because we consider the top Yukawa interaction, as we will see shortly.
We collectively denote the gauge interactions as $a = 1, 2$, and $3$
for U(1)$_Y$, SU(2)$_L$, and SU(3)$_c$, respectively.
As noted above, we denote the primary particle injected from a heavy particle decay as $\si$.

Here, we summarize group factors for later convenience.
We use F and A to denote fundamental and adjoint representations.
In general, we have the following equalities, $t_{\rm R} = d_{\rm R} C_{\rm R}/ d_{\rm A}$, and $\ta = \ca$.
For SU($N$),
$C_{\rm F}^{(N)} = (N^2-1)/(2N)$, $C_{\rm A}^{(N)}= N$,
$d_{\rm F}^{(N)} = N$, $d_{\rm A}^{(N)} = N^2-1$, $t_{\rm F}^{(N)} = 1/2$, and $t_{\rm A}^{(N)} = N$.
For U(1)$_Y$,
$d_{\rm F}^{(1)} = d_{\rm A}^{(1)} = 1$, and
$C_{{\rm R}_s}^{(1)} = t_{{\rm R}_s}^{(1)} = q_{Y,s}^2$ for a particle $s$,
and $C_{\rm A}^{(1)} = 0$ for gauge bosons.
Explicitly,
\begin{align}
   C_{{\rm F}}^{(3)} = \tfrac43 \,,
   \qquad
   C_{{\rm A}}^{(3)} = 3 ,
   \qquad
   \df^{(3)} = 3 ,
   \qquad
   d_{{\rm A}}^{(3)} = 8 ,
   \qquad
   \tf^{(3)} = \tfrac12 ,
   \qquad
   t_{{\rm A}}^{(3)} = 3 ,
   \\
   C_{{\rm F}}^{(2)} = \tfrac34 \,,
   \qquad
   C_{{\rm A}}^{(2)} = 2 ,
   \qquad
   d_{\rm F}^{(2)} = 2 ,
   \qquad
   d_{{\rm A}}^{(2)} = 3 ,
   \qquad
   \tf^{(2)} = \tfrac12 ,
   \qquad
   \ta^{(2)} = 2 ,
   \\
   C_{{\rm F}_s}^{(1)} = q_{Y,s}^2 \,,
   \quad
   C_{{\rm A}}^{(1)} = 0 ,
   \qquad
   \df^{(1)} = 1 ,
   \qquad
   d_{{\rm A}}^{(1)} = 1 ,
   \qquad
   t_{{\rm F}_s}^{(1)} = q_{Y,s}^2,
   \quad
   t_{{\rm A}}^{(1)} = 0,
\end{align}
for charged particles.
If a particle $s$ is not charged under $a$,
we define $C_{{\rm R}_s}^{(a)} = 0$, $d_{{\rm R}_s}^{(a)} = 1$, and $t_{{\rm R}_s}^{(a)} = 1$.

We define $\alpha_t \equiv y_t^2 / (4\pi)$,
where $y_t$ represents the top Yukawa coupling.
We neglect Yukawa interactions other than the top Yukawa interaction due to its smallness.
In our numerical calculation, we use
\begin{align}
 &\alpha_1 (m_Z) = (1-\sin^2\theta_W)^{-1} \alpha(m_Z) \simeq 0.0102
 \qquad
 \alpha_2 (m_Z) = \sin^{-2} \theta_W \alpha(m_Z) \simeq 0.0338
 \\
 &\alpha_3 (m_Z) \simeq  0.118
 \qquad \qquad \qquad \qquad \qquad \qquad
 \alpha_t (m_t) \simeq  0.0786
\end{align}
where $\theta_W$ ($\sin^2 \theta_W \simeq 0.231$) is the Weinberg angle, $\alpha (m_Z)$ ($\simeq 1/128$) is the fine-structure constant, and
$m_Z$ ($\simeq 91.2 \GeV$) and $m_t$ ($\simeq 173 \GeV$) are the $Z$-boson and top masses, respectively~\cite{Workman:2022ynf}.
The gauge coupling runs, such as
\begin{align}
 \alpha_a^{-1} (\mu) -  \alpha_a^{-1} (\mu_0)
&= - \frac{b_a}{4 \pi} \ln \frac{\mu^2}{\mu_0^2},
 \\
 b_a &= \frac{4}{3} \sum_s \frac{t_{{\rm R}_s}^{(a)} \nu_s}{4 d_{{\rm R}_s}^{(a)}} B_s - \frac{11}{3} C_{\rm A}^{(a)}
 \\
 &=\left\{
   \begin{array}{lll}
   -7 \quad \text{for \ SU(3)}
   \\
   -\frac{19}{6}  \quad \text{for \ SU(2)}
   \\
   \frac{41}{6}  \quad \text{for \ U(1)}_Y
   \end{array}
   \right. ,
\end{align}
where $B_s = 1$ for a fermion and $1/2$ for a boson.
We also use
\begin{align}
 &\alpha_t^{-1} (\mu) -  \alpha_t^{-1} (\mu_0)
= - \frac{b_t}{4 \pi} \ln \frac{\mu^2}{\mu_0^2},
 \\
 &b_t
 = \alpha_t^{-1}
 \qty( \frac{9}{2} \alpha_t
 -8 \alpha_3
 - \frac{9}{4} \alpha_2
 - \frac{17}{12} \alpha_1
 )
\end{align}
for the running of the top Yukawa coupling.
We are mainly interested in thermalization of SM particles with an energy scale much larger than the electroweak scale.
Because of the renormalization group running, the gauge coupling constants are of the same order with each other at such a high energy scale. We thus perform a resummation for all gauge interactions on an equal footing to calculate $\mu_\perp^2$ [see Eq.~\eqref{eq:mu_perp}]~\cite{Anisimov:2010gy,Besak:2012qm}.

%%%%%%%%%%%%%%%%%%%%%%%%%%%%%%%%%%%%%%%%%%%%%%%%%%%%%%%%%%%%%%%%%%%%%%%%%%%%%%%%%%%%%%%%%%%%%%%%%%%%

We define the distribution function $f_s$ for each species.
It is normalized such that
$f(p) = 1/(e^{p/T} \mp 1)$ without the degrees of freedom $\nu_s$ in thermal equilibrium for bosons and fermions.
We are interested in the case with an under-dense regime, where $f_s(p) \ll 1$ for hard particles.
We also consider the case with $p_0 \gg T$, which allows us to approximate the splitting functions
using the next-to-leading-logarithm approximation.
In this case, the Boltzmann equation is reduced to
\begin{align}
  \frac{\partial }{\partial t} f_s (p,t)
  &= - \frac{(2\pi)^3}{p^2 \nu_s}  \sum_{s',s''}
  \int_0^p  \dd k \,
    \gamma_{s \leftrightarrow s's''} \bigl(p; k, p-k \bigr) \,
    f_s(p)
    +
    \frac{(2\pi)^3}{p^2 \nu_s} \sum_{s',s''}
  \int_0^\infty \dd k \,
    \gamma_{s' \leftrightarrow s s''} \bigl(p+k; p, k \bigr) \,
    f_{s'}(p+k)
    \nn
    &\qquad + ({\rm source \ term}),
\label {eq:boltzmann}
\end{align}
where the final line represents the source term.
Summation over $s'$ and $s''$ is taken for all particle species multiplied by the number of flavors in the unit of Weyl fermions and complex scalar fields.
The explicit forms of the splitting rate $\gamma_{s \leftrightarrow  s' s''}$ and Boltzmann equations are written in the next section and Appendix~\ref{sec:appendixA}.

%%%%%%%%%%%%%
\subsection{Splitting rate for the SM}
\label{sec:SM}
%%%%%%%%%%%%%

The splitting functions $\gamma_{s \leftrightarrow  s' s''}\bigl(P; xP, (1-x)P\bigr)$ include the summation over the spin degrees of freedom of a chiral fermion and a complex scalar field with a single flavor (or one-half of a Dirac field) with respect to the relevant gauge group.
The next-to-leading-logarithm result can be summarized in the following
form~\cite{Arnold:2001ba,Arnold:2002ja,Arnold:2002zm,Anisimov:2010gy,Bodeker:2019ajh}:
\begin {subequations}
\label{eq:gammas}
\begin{align}
   \gamma_{g_a\leftrightarrow g_ag_a}(P; xP, (1-x)P)
   &= \frac{1}{2} \frac{d_{{\rm A}}^{(a)} C_{{\rm A}}^{(a)} \alpha_a}{(2\pi)^4 \sqrt2}
      \,
     \frac{1^4+x^4+(1-x)^4}{1^2 \cdot x^2(1-x)^2}\, \mu_{\perp,a}^2(1,x,1{-}x;\, g_a,\,g_a,\, g_a) \, ,
\label {eq:gamma_ggg}
\\
   \gamma_{s \leftrightarrow g_a s}(P; xP, (1-x)P)
   &= \frac12 \frac{d_{{\rm F}}^{(a)} C_{{\rm F}_s}^{(a)} \alpha_a}{(2\pi)^4 \sqrt2}
      \,
     \frac{1^2+(1-x)^2}{1 \cdot x^2(1-x)}\, \mu_{\perp}^2(1,x,1{-}x;\, s,g_a,s)
     \quad {\rm for} \ s = {\rm (fermion)}\,,
\label {eq:gamma_qgq}
\\
   \gamma_{g_a \leftrightarrow s\bar s}(P; xP, (1-x)P)
   &= \frac12 \frac{d_{{\rm F}}^{(a)} C_{{\rm F}_s}^{(a)} \alpha_a}{(2\pi)^4 \sqrt2}
      \,
     \frac{x^2+(1-x)^2}{1^2 \cdot x(1-x)} \, \mu_{\perp}^2(1,x,1{-}x;\, g_a,s,s)
     \quad {\rm for} \ s = {\rm (fermion)}\,,
\label {eq:gamma_gqq}
\\
   \gamma_{\phi \leftrightarrow g_a \phi}(P; xP, (1-x)P)
   &= \frac12 \frac{d_{{\rm F}}^{(a)} C_{{\rm F}_\phi}^{(a)} \alpha_a}{(2\pi)^4 \sqrt2}
      \,
     \frac{2}{x^2}\, \mu_{\perp}^2(1,x,1{-}x;\, \phi,g_a,\phi) \,,
\label {eq:gamma_pgp}
\\
   \gamma_{g_a \leftrightarrow \phi \phi^*}(P; xP, (1-x)P)
   &= \frac12 \frac{d_{{\rm F}}^{(a)} C_{{\rm F}_\phi}^{(a)} \alpha_a}{(2\pi)^4 \sqrt2}
      \,
     \frac{2}{1^2} \, \mu_{\perp}^2(1,x,1{-}x;\, g_a,\phi,\phi) \,,
\label {eq:gamma_gpp}
\\
   \gamma_{u_3 \leftrightarrow \phi Q_3}(P; xP, (1-x)P)
   &= \frac12 \frac{\alpha_{\rm t}}{(2\pi)^4 \sqrt2}
      \,
     \frac{1}{1 \cdot (1-x)}\, \mu_{\perp}^2(1,x,1{-}x;\, {u_3},\phi,{Q_3}) \,,
\label {eq:gamma_qpq}
\\
   \gamma_{\phi \leftrightarrow u_3 \bar{Q}_3}(P; xP, (1-x)P)
   &= \frac12 \frac{\alpha_{\rm t}}{(2\pi)^4 \sqrt2}
      \,
     \frac{1}{x(1-x)} \, \mu_{\perp}^2(1,x,1{-}x;\, \phi,{u_3},{Q_3}) \,,
\label{eq:gamma_pqq}
\end{align}
\end {subequations}
where $g_a$ collectively represent the gauge bosons of gauge group $G_a$.%
\footnote{
We add a factor of $1/2$ to \eqref{eq:gamma_ggg} as a symmetry factor to avoid a double count,
where we integrate over $x$ from $0$ to $1$ rather than from $0$ to $1/2$ in the Boltzmann equation.
}
Note that we add a factor of $1/2$ to Eqs.~(\ref{eq:gamma_qgq}), (\ref{eq:gamma_gqq}), (\ref{eq:gamma_qpq}), and (\ref{eq:gamma_pqq}) because we use chiral fermions rather than a Dirac fermion.
Contributions from switched diagrams are included in the splitting functions.
For example,
the rates of processes such as $\bar{q} \leftrightarrow g_a \bar{q}$ are included in
$\gamma_{q \leftrightarrow g_a q}$, where $\bar{q}$ represents the anti-particle of $q$.
If we want to treat
$u_3 \leftrightarrow \phi Q_3$
and
$Q_3 \leftrightarrow \phi^* u_3$
separately,
we should multiply $\gamma_{u_3 \leftrightarrow \phi Q_3}$ ($= \gamma_{Q_3 \leftrightarrow \phi^* u_3}$) by a factor of $1/2$ for each splitting rate [see Eqs.~(\ref{eq:f_u3}), (\ref{eq:f_Q3}), and (\ref{eq:f_phi})].

The functions of $x$ in the above equations come from the DGLAP splitting functions.
The function $\mu_\perp^2$ is given by \eqref{eq:mu_perp}
with $G=$SU(3)$_c\times$SU(2)$_L\times$U(1)$_Y$,
where
\begin{align}
   {\cal N}_a
  &=\left\{
   \begin{array}{lll}
   15 \frac{\zeta(3)}{\pi^2} T^3 \quad \text{for \ SU(3)}
   \\
   14 \frac{\zeta(3)}{\pi^2} T^3  \quad \text{for \ SU(2)}
   \\
   6 \frac{\zeta(3)}{\pi^2} T^3  \quad \text{for \ U(1)}_Y
   \end{array}
   \right.
\\
   m_{D,a}^2
   &=\left\{
   \begin{array}{lll}
   8 \pi \alpha_3 T^2 \quad \text{for \ SU(3)}
   \\
   \frac{22 \pi }{3} \alpha_2 T^2  \quad \text{for \ SU(2)}
   \\
   \frac{22 \pi}{3} \alpha_1 T^2  \quad \text{for \ U(1)}_Y
   \end{array}
   \right.
\end{align}
from Eqs.~(\ref{eq:N_a}) and (\ref{eq:mD}).

The Boltzmann equations of the SM are as follows: The explicit form of each species is shown in Appendix~\ref{sec:appendixA}.
For a gauge boson $g_a$,
\begin{align}
  \text{(splitting terms)}
  &= - \frac{(2\pi)^3}{p^2 \nu_{g_a}}
  \int_0^p  \dd k \,
    \qty[ \gamma_{g_a \leftrightarrow g_a g_a} \bigl(p; k, p-k \bigr) + \sum_{s'} \frac{d_{{\rm R}_{s'}}^{(2)}d_{{\rm R}_{s'}}^{(3)}}{d_{{\rm R}_{s'}}^{(a)}} \gamma_{g_a \leftrightarrow s' \bar{s}'} \bigl(p; k, p-k \bigr)
    ]
    f_{g_a}(p)
    \nn
    &+
    \frac{(2\pi)^3}{p^2 \nu_{g_a}}
  \int_0^\infty \dd k \,
    \qty[
    2 \gamma_{g_a \leftrightarrow g_ag_a} \bigl(p+k; p, k \bigr) f_{g_a}(p+k)
    + \sum_{s'} \frac{d_{{\rm R}_{s'}}^{(2)}d_{{\rm R}_{s'}}^{(3)}}{d_{{\rm R}_{s'}}^{(a)}} \gamma_{s' \leftrightarrow g_a s'} \bigl(p+k; p, k \bigr) f_{s'}(p+k)
    ]
    .
\end{align}
where $s'$ represents the fermions and Higgs.
The summation over $s'$ should include the contributions from all flavors.
Note that the self-splitting process is absent for the Abelian gauge boson.
For a fermion or scalar $s$,
\begin{align}
  \text{(splitting terms)}
  &= - \frac{(2\pi)^3}{p^2 \nu_{s}}
  \int_0^p  \dd k \,
    \sum_a \frac{ d_{{\rm R}_s}^{(2)}d_{{\rm R}_s}^{(3)}}{d_{{\rm R}_s}^{(a)}} \gamma_{s \leftrightarrow g_a s} \bigl(p; k, p-k \bigr)
    f_{s} (p)
    \nn
    &+
    \frac{(2\pi)^3}{p^2 \nu_{s}}
  \int_0^\infty \dd k \,
    \qty[ \sum_a \frac{ d_{{\rm R}_s}^{(2)}d_{{\rm R}_s}^{(3)}}{d_{{\rm R}_s}^{(a)}}
    \qty(
     2 \gamma_{g_a \leftrightarrow s \bar{s}} \bigl(p+k; p, k \bigr) f_{g_a}(p+k)
    + \gamma_{s \leftrightarrow s g_a} \bigl(p+k; p, k \bigr) f_{s}(p+k) )
    ]
    .
    \label{eq:boltzmannparticle}
\end{align}
where we omit Yukawa interactions.
The contribution from the top Yukawa interaction is given by
\begin{align}
  \text{(splitting terms)} \ni
   &- \frac{(2\pi)^3}{p^2 \nu_{s}}
  \int_0^p  \dd k \,
     d_{{\rm R}_s}^{(3)} \gamma_{s \leftrightarrow \phi s'} \bigl(p; k, p-k \bigr)
    f_{s} (p)
    \nn
    &+
    \frac{(2\pi)^3}{p^2 \nu_{s}}
  \int_0^\infty \dd k \,
    d_{{\rm R}_s}^{(3)}
    \qty[
     2 \gamma_{\phi \leftrightarrow s s'} \bigl(p+k; p, k \bigr) f_\phi (p+k)
    + \gamma_{s' \leftrightarrow s \phi} \bigl(p+k; p, k \bigr) f_{s'}(p+k)
    ]
    .
\end{align}
for the top quark $s = u_3$ and $Q_3$ with $s' = Q_3$ and $u_3$, respectively,
and
\begin{align}
  \text{(splitting terms)} \ni
   &- \frac{(2\pi)^3}{p^2 \nu_{\phi}}
  \int_0^p  \dd k \,
     2 d_{{\rm R}_s}^{(3)} \gamma_{\phi \leftrightarrow s s'} \bigl(p; k, p-k \bigr)
    f_{\phi} (p)
    \nn
    &+
    \frac{(2\pi)^3}{p^2 \nu_{\phi}}
  \int_0^\infty \dd k \,
    d_{{\rm R}_s}^{(3)}
    \qty[
     \gamma_{s \leftrightarrow \phi s'} \bigl(p+k; p, k \bigr) f_s (p+k)
    + \gamma_{s' \leftrightarrow \phi s} \bigl(p+k; p, k \bigr) f_{s'}(p+k)
    ] .
\end{align}
for the Higgs $\phi$.

We note that the Boltzmann equations are linear in terms of the distributions so that the normalization of distributions can be arbitrary. This in turn implies that we can always rescale $\tilde{\Gamma}$ without loss of generality.
We thus take $\tilde{\Gamma} = 1$ in our numerical calculations for any $\si$ in Sec.~\ref{sec:numerical}.
The linearity of the equations also implies that we can take $\si$ independently for all species.
Moreover, we can take a linear combination of $\si$ to obtain the solution with an arbitrary primary source term.
Our results for $\si = \qty( e_f, L_f, u_{f'}, u_3, d_f, Q_{f'}, Q_3, \phi, g, W, B )$ therefore cover all possible initial conditions for the source term.
When we consider the case in which the injected particle is solely for $\si$ with a particular flavor,
the branching rate $\Br$ is given by
\begin{align}
    \Br = \frac{1}{\nu_s} \delta_{s\, \si},
    \label{eq:Br}
\end{align}
where $\delta_{s\, \si}$ is the Kronecker delta.
If we equally treat all flavors (except for the top quarks)
and assume that high-energy particles are injected into all $f$ or $f'$,
the branching ratio should be multiplied by $1/3$ or $1/2$, respectively.

%%%%%%%%%%%%%
\section{Analytic calculations}
\label{sec:analytic}
%%%%%%%%%%%%%

Before we solve the Boltzmann equations numerically, in this section, we provide analytic results at $p \approx p_0$ and $p \ll p_0$.
The asymptotic behavior at $p \approx p_0$ provides an appropriate boundary condition for the distribution function at $p = p_0$ from the delta-function source term.
This is useful for the numerical calculations.
The asymptotic behavior at $p \ll p_0$ is phenomenologically important for discussing non-thermal DM production during the thermalization process of SM particles.
We will see that these analytic results are consistent with our numerical results in Sec.~\ref{sec:numerical}.

\subsection{Boundary condition and asymptotic behavior at $p \approx p_0$}

\begin{figure}[t]
	\centering
 	\includegraphics[width=0.5\linewidth]{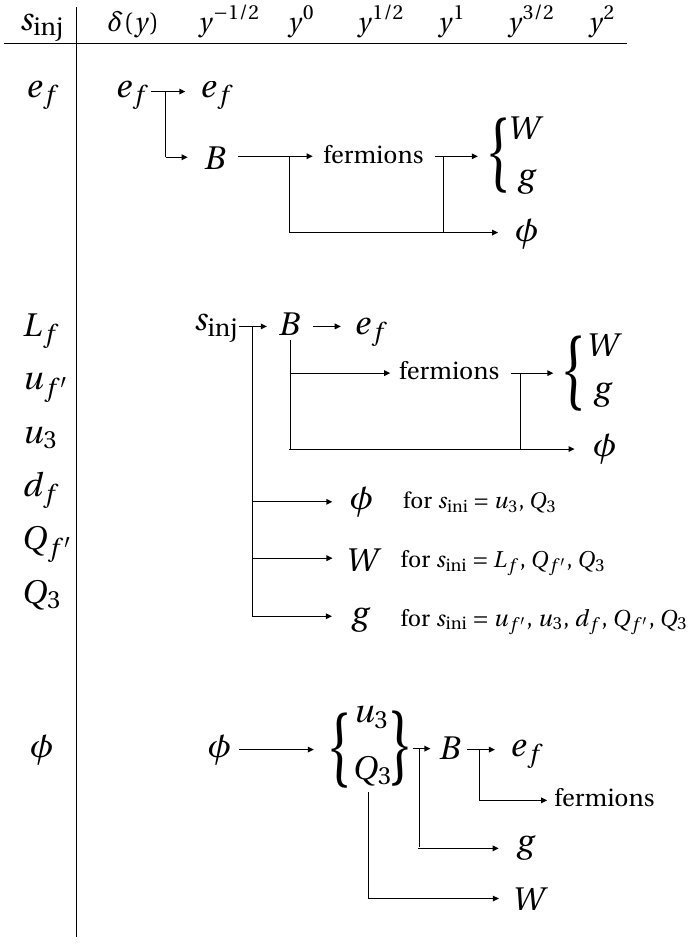}
 	\hspace{0.4cm}
 	\includegraphics[width=0.45\linewidth]{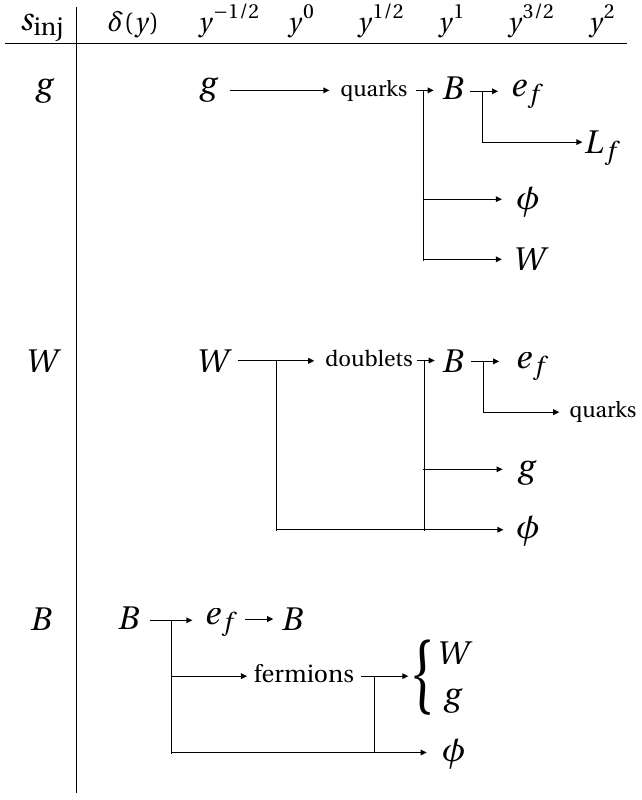}
	\caption{
	Cascades of primary particles. The leftmost part represents the primary particles $\si$. Particle production occurs from left to right, as indicated by the arrows.
	The horizontal axis represents the behavior of distributions at $p \approx p_0$, indicated by the function of $y \equiv (p_0 - p)/p_0$ ($\ll 1$) at the first line. ``fermions'' represents the fermions that are not already produced in the cascades.
	For the case of $\si = e_f$ and $B$, the distribution of primary particles is a delta function plus some function of $y$. For the other cases, the distribution of primary particles is proportional to $y^{-1/2}$ without $\delta(y)$.
    }
	\label{fig:flow1}
\end{figure}

Here, we explain the asymptotic behaviors of distributions at $p \approx p_0$, which are useful for imposing boundary conditions on distributions in numerical calculations.
The asymptotic behavior is different for the primary particle and other particles.
The splitting process at $p\approx p_0$ is schematically represented in Fig.~\ref{fig:flow1},
where the leftmost part represents the primary particle $\si$.
The function of $y \equiv (p_0 - p)/p_0$ at the top of the figure represents the behavior of the distribution of the corresponding species at $p \approx p_0$.
The arrows represent the cascades of particles from primary particles. In this section, we provide summary results for the distributions of only primary particles. Their derivation and the secondary and subsequent distributions are given in Appendix~\ref{sec:appendixB} and are schematically summarized in Fig.~\ref{fig:flow1}.

First, let us consider a case in which gluons are injected at $p = p_0$ from a delta-function source term.
The stationary equation for the gluon distribution can be expressed as
\begin{align}
  - 2 \int_0^{p/2}  \dd k \,
    \gamma_{g \leftrightarrow g g} \bigl(p; k, p-k \bigr) \,
    f_g(p)
    +
  \int_0^\infty \dd k \,
    2 \gamma_{g \leftrightarrow g g} \bigl(p+k; p, k \bigr) \,
    f_g(p+k)
    + \frac{p^2}{(2\pi)^3} p_0^{1/2} T^{3/2} \tilde{\Gamma} \delta (p - p_0) = 0,
\end{align}
where we include the source term ($= p_0^{1/2} T^{3/2} \Br \tilde{\Gamma} \delta (p - p_0)$) and use $\Br = 1/\nu_g$ [see Eq.~\eqref{eq:Br}].
Here, we neglect terms associated with quarks because the secondary particles are subdominant at $p \approx p_0$ and there is no IR divergence in the pair production of quarks.

In Appendix~\ref{sec:appendixB},
we discuss that the following distribution function satisfies the above equation:
\begin{equation}
 f_g(p) \approx \frac{\tilde{\Gamma}}{ (2\pi)^4 \tilde{\gamma}_{g \leftrightarrow g g}}  \qty( \frac{p_0 - p}{p_0} )^{-1/2}
\end{equation}
for $p \approx p_0$, where $\tilde{\gamma}_{g \leftrightarrow g g}$ is defined below \eqref{eq:gluonasym}.
This represents the asymptotic behavior of the gluon when the primary particles are gluons, $\si = g$.

Similar results hold for $\si = L_f, u_f, d_f, Q_f, \phi$, and $W$ once we replace $\tilde{\gamma}_{g \leftrightarrow g g}$, as appropriate.
Below, we summarize their results.
\begin{equation}
 f_{s_1} (p) \approx
 \frac{\tilde{\Gamma}}{(2\pi)^4 \sum_{a'} \tilde{\gamma}_{s_{1} \to g_{a'} s_{1}}
 }  \qty( \frac{p_0 - p}{p_0} )^{-1/2}
\end{equation}
for $s_1 = L_f, u_{f'}, u_3, d_f, Q_{f'}, Q_3, \phi, g$, and $W$,
where $\tilde{\gamma}$ are defined in Eqs.~(\ref{eq:tildegtogg}), (\ref{eq:tildeqtogq}), and (\ref{eq:tildephitogphi}).
As a reference, we obtain
\begin{align}
(2\pi)^4 \sum_{a'} \tilde{\gamma}_{s_{1} \to g_{a'} s_{1}}
 &\simeq
 (
 0.053, \
 0.23, \
 0.23, \
 0.23, \
 0.62, \
 0.62, \
 0.053, \
 4.9, \
 0.42
 )
 \nonumber\\
 &{\rm for} \ s_1 = (L_f, \ u_{f'}, \ u_3, \ d_f, \ Q_{f'}, \ Q_3, \ \phi, \ g, \ W),
 \label{eq:gammatilde1}
\end{align}
respectively.
Here, we assume $p = p_0 = 10^{12} T = 10^{15} \GeV$, though the dependence on $p$ is only logarithmic through the running of coupling constants.
These $\tilde{\gamma}$ represent the splitting rates of the primary particles at $p \approx p_0$.

The cases with $\si = e_f$ and $B$ have qualitatively different distributions because they experience only Abelian gauge interactions leading to a different scaling for the splitting rate [see Eq.~\eqref{eq:split_func_rough}].
If the primary particle is an Abelian gauge boson $B$, it splits only into other particles and does not undergo soft-dominated splitting processes.
In this case, we obtain
the delta-function distribution for $f_B(p)$ as
\begin{equation}
 f_B(p) = C_B' \delta (p - p_0) + C_B \qty( \frac{p_0 - p}{p_0} )^{0},
\end{equation}
where we include
a subdominant component, which is determined in Appendix~\ref{sec:appendixB}.
Here, $C_B'$ is given by
\begin{equation}
 C_B'
    = \frac{p_0}{(\pi/4) (2\pi)^3} \frac{\tilde{\Gamma}}{3 \sum_{s_f} \tilde{\gamma}_{B \to s_f s_f} +  \tilde{\gamma}_{B \to \phi \phi} }.
\end{equation}
where $s_f$ represents fermions.
As a reference, we obtain
\begin{align}
(\pi/4)(2\pi)^3 \qty[ \sum_{s_f} \tilde{\gamma}_{B \to s_f s_f} +  \tilde{\gamma}_{B \to \phi \phi} ]
 &\simeq
 0.041,
 \label{eq:gammatilde2}
\end{align}
where we assume $p = p_0 = 10^{12} T = 10^{15} \GeV$, though the dependence on $p$ is only logarithmic.

If a right-handed lepton is produced from heavy particle decay, that is, if $\si = e_f$,
the soft-photon emission of $e_f \to B e_f$ is more suppressed by the LPM effect than the non-Abelian case and again cannot soften the delta-function distribution.
As a result, it remains a delta-function distribution plus a subdominant component such as
\begin{equation}
 f_{e_f} (p) = C_{e_f}' \delta (p - p_0) + C_{e_f} \qty( \frac{p_0 - p}{p_0} )^{-1/2}.
\end{equation}
Here,
$C_{e_f}'$ is given by
\begin{align}
  C_{e_f}'
  \approx \frac{p_0 }{(2\pi)^3 }
  \frac{8}{11 \pi} \frac{ \tilde{\Gamma}}{\tilde{\gamma}_{e_f \to B e_f}}.
\end{align}
As a reference, we obtain
\begin{align}
\frac{11 \pi}{8} (2\pi)^3 \tilde{\gamma}_{e_f \to B e_f}
 &\simeq
 1.3 \times 10^{-3},
 \label{eq:gammatilde3}
\end{align}
where we assume $p = p_0 = 10^{12} T = 10^{15} \GeV$, though the dependence on $p$ is only logarithmic.

Other particles are produced from the splitting of primary particles.
Secondary and subsequent particles have suppressed distributions at $p \approx p_0$ with some power of $(p_0 - p)/p_0$. The power of this behavior can be determined analytically, as discussed in Appendix~\ref{sec:appendixB}. The result is summarized in Fig.~\ref{fig:flow1}.
The leftmost part represents the primary particle $\si$.
The arrows represent the cascades from the primary particle, namely, they demonstrate how secondary and subsequent particles are produced from the primary particle.
The behavior of the distributions of corresponding particles is represented by the uppermost line, where $y \equiv (p_0 - p)/p_0$.
If the primary particle is $e_f$ or $B$,
they have the delta-function distribution $\delta (y)$, as discussed above.
The other primary particles have a distribution proportional to $y^{-1/2}$.
For example,
if the primary particle is a gluon ($\si = g$),
its distribution is $\propto y^{-1/2}$.
The secondary particles produced by the splitting from the gluon are quarks,
which have the distribution $\propto y^{1/2}$.
The U(1)$_Y$ gauge boson $B$ as well as the Higgs $\phi$ and W bosons are produced from the splitting of quarks, where the distribution is $\propto y^1$ for $B$ and $y^{3/2}$ for $\phi$ and $W$. Right- and left-handed leptons are produced from the splitting of $B$, and the distribution is $\propto y^{3/2}$ and $y^2$, respectively.
As we will explain shortly,
we confirm the behaviors described in Fig.~\ref{fig:flow1} using numerical calculations.

\subsection{Asymptotic behavior at $p \ll p_0$}
\label{sec:analyticsmall}

It is known that the splitting process results in
$f \propto p^{-7/2}$ for $p \ll p_0$ (see, e.g., Refs~\cite{Kurkela:2011ti,Kurkela:2014tea}).
As we will see in Sec.~\ref{sec:numerical}, this behavior is confirmed by our numerical results, even for a case with entirely SM particles.
In this section, we first derive the ratio of the distributions for different species at $p \ll p_0$.
For this purpose, we define $R_g^{(s)}(p)$ as the fraction of the number of species $s$ at energy $p$.
\begin{equation}
 R_s(p) \equiv
 \frac{\nu_s f_s(p)}{\sum_{s'} \nu_{s'} f_{s'}(p)},
 \label{eq:Rs}
\end{equation}
where the summation in the denominator is taken for all families and species.
We expect that they reach the constant values $R_s^{\rm (asym)}$ at $p \ll p_0$.

We can derive the constraint equations that determine $R_s^{\rm (asym)}$ by substituting $f_s (p) = \tilde{f}_s p^{-7/2}$ into the Boltzmann equations, where $\tilde{f}_s$ are constants, and performing integrations.
The constraint equations are given by
\begin{align}
 &- 0.0354 \tilde{f}_B + 0.0265 \tilde{f}_{e_f} = 0,
 \label{eq:constrainte}
 \\
 &- 0.0370 \tilde{f}_B +
 0.861 \tilde{f}_{L_f} - 0.285 \tilde{f}_W = 0,
\label{eq:constraintL} \\
 &- 0.179 \tilde{f}_B - 1.71 \tilde{f}_g +
 7.15 \tilde{f}_{u_{f'}} = 0,
 \\
 &- 0.179 \tilde{f}_B - 1.71 \tilde{f}_g - 3.38 \tilde{f}_\phi -
 0.231 \tilde{f}_{Q_3} + 8.52 \tilde{f}_{u_3} = 0,
 \\
 &- 0.0447 \tilde{f}_B + 7.05 \tilde{f}_{d_f} -
 1.70 \tilde{f}_g = 0,
 \\
 &-0.0119 \tilde{f}_B - 1.96 \tilde{f}_g + 8.30 \tilde{f}_{Q_{f'}} -
 0.964 \tilde{f}_W = 0,
 \\
 & -0.0119 \tilde{f}_B - 1.96 \tilde{f}_g - 1.64 \tilde{f}_\phi +
 8.93 \tilde{f}_{Q_3} - 0.130 \tilde{f}_{u_3} - 0.964 \tilde{f}_W = 0,
 \\
 &-
 0.0123 \tilde{f}_B + 10.7 \tilde{f}_\phi - 1.54 \tilde{f}_{Q_3} - 1.67 \tilde{f}_{u_3} -
 0.0796 \tilde{f}_W = 0,
 \\
 &- 7.89 \tilde{f}_B + 8.25 \tilde{f}_g - 10.6 \tilde{f}_{Q_{f'}} -
 5.31 \tilde{f}_{Q_3} - 2.63 \tilde{f}_{u_3} - 5.26 \tilde{f}_{u_{f'}} = 0,
 \\
 &- 1.67 \tilde{f}_{L_f} -
 0.434 \tilde{f}_\phi - 4.82 \tilde{f}_{Q_{f'}} - 2.41 \tilde{f}_{Q_3} + 6.41 \tilde{f}_W = 0,
 \\
 &2.58 \tilde{f}_B - 0.301 \tilde{f}_{d_f} - 0.0796 \tilde{f}_{e_f} - 0.167 \tilde{f}_{L_f} -
 0.0247 \tilde{f}_\phi - 0.107 \tilde{f}_{Q_{f'}} - 0.0535 \tilde{f}_{Q_3} -
 0.402 \tilde{f}_{u_3} - 0.805 \tilde{f}_{u_{f'}} = 0
 \label{eq:constraintgamma}
\end{align}
for $s = e_f, L_f, u_{f'}, u_3, d_f, Q_{f'}, Q_3, \phi, g, W$, and $B$, respectively,
where we multiply all terms by $10^3$.
These constraint equations cannot be solved exactly because
the Boltzmann equation includes running gauge coupling, which depends logarithmically on $p$.
However, we can still determine an approximate value of $\tilde{f}_s$.
For example, we can solve Eqs.~(\ref{eq:constraintL}-\ref{eq:constraintgamma})
without imposing \eqref{eq:constrainte}
and check how accurately \eqref{eq:constrainte} is satisfied.
We find that the error on \eqref{eq:constrainte} is of the order of $3 \times 10^{-6}$, which means that all constraint equations are satisfied with errors of $\mathcal{O}(0.01\%)$ at most.
We can calculate $R_s^{\rm (asym)}$ from
the constraint equation results
such as
\begin{equation}
 R_s^{\rm (asym)} =
 \frac{\nu_s \tilde{f}_s}{\sum_s \nu_s \tilde{f}_s},
 \label{eq:Rs2}
\end{equation}
where the factor of $p^{-7/2}$ is cancelled between the numerator and denominator.
The result is shown in Tab.~\ref{tab:1}.
We find that the gluon dominates in terms of the number and energy of SM particles in the scaling regime.
These asymptotic behaviors are verified using our numerical calculations, as shown in Sec.~\ref{sec:numerical},
where
we also determine the energy scale $p = p_{\rm asym}^{(s_{\rm inj})}$ below which $R_s(p) \simeq R_s^{\rm (asym)}$ from numerical calculations.

We also define
an asymptotic value of total distributions such as
\begin{align}
 f_{\rm tot}^{\rm (asym)} (p)
 &\equiv
 \left. \sum_s \nu_s f_s(p) \right\vert_{p < p_{\rm asym}^{(s_{\rm inj})}},
 \label{eq:totaldist}
 \\
 &\equiv
 \tilde{\Gamma} \tilde{f}_{\rm tot}^{\rm (asym)}
 \qty( \frac{p}{p_0} )^{-7/2}.
 \label{eq:tildetotal}
\end{align}
This is proportional to $(p/p_0)^{-7/2}$ and $\tilde{\Gamma}$, which we explicitly show in Eq.~(\ref{eq:tildetotal}).
We expect that $\tilde{f}_{\rm tot}^{\rm (asym)}$ is independent of the injected particle $\si$ because of the following reason.
We first note that
the fraction of distributions, $R_s^{(\rm asym)}$,
is independent of $\si$, as shown above.
We then note that the source term continuously provides a constant energy per unit time. The injected energy is transferred to a smaller momentum $p$ via the splitting. Because of the conservation of energy per unit time, we expect that the energy transfer in the distribution from a large momentum to a small momentum depends only on the injected energy and does not depend on the injected particle species $\si$.
This implies that the asymptotic values of distributions are determined by how fast the thermalization process occurs at a small $p$.
Because the fraction of distributions is also independent of $\si$,
the amplitude of the total distribution function is then expected to be independent of the injected primary particles.
This is actually confirmed by our numerical calculations for all $\si$ as we will see in Sec.~\ref{sec:numerical}.

Here we provide a rough estimation~\cite{Harigaya:2014waa,Harigaya:2019tzu}.
Focusing on a distribution around a momentum $p$ ($\ll p_0$),
the time scale for the energy flow out of that momentum scale is given by $\left. \Gamma_{\rm LPM}^{-1}\right\vert_{k \sim p}$, which implies the energy flow out of that momentum scale is estimated as
\begin{align}
 p^4 \left. \Gamma_{\rm LPM}\right\vert_{k \sim p} f_{\rm tot}^{(\rm asym)}(p)
 \sim \alpha \alpha_d T^{3/2} p^{7/2} \tilde{\Gamma} \tilde{f}_{\rm tot}^{\rm (asym)} \qty( \frac{p}{p_0} )^{-7/2}
\end{align}
where we use Eq.~\eqref{eq:split_func_rough}.
This should be compared with the energy injection per unit time:
\begin{align}
 \int 4 \pi p^3 \dd p  \, 2 \Gamma_I \delta ( p - p_0) \frac{2\pi^2}{p^2} n_I =
 4 \pi p_0^2 \tilde{\Gamma} p_0^{3/2} T^{3/2}.
\end{align}
We then obtain $\tilde{f}_{\rm tot}^{\rm (asym)} \sim 1/\alpha \alpha_d$, where we omit numerical factors for simplicity.
Because the asymptotic behavior is determined by the thermalization or splitting process of the dominant component, gluons,
we finally obtain
\begin{align}
 \tilde{f}_{\rm tot}^{\rm (asym)} \sim \alpha_s^{-2}.
 \label{eq:tildeftotest}
\end{align}

Combining Eqs.~\eqref{eq:Rs2} and \eqref{eq:tildetotal}, we obtain
\begin{equation}
 \left. \nu_s f_s(p) \right\vert_{p < p_{\rm asym}^{(s_{\rm inj})}}  = \tilde{\Gamma} R_s^{\rm (asym)}
 \tilde{f}_{\rm tot}^{\rm (asym)} \qty( \frac{p}{p_0} )^{-7/2}.
\end{equation}
These results indicate that any particles can be produced from any primary particles through the cascades.
Moreover, even the fraction of particles is the same for any primary particles.
This is important if one considers DM production through the collision of these secondary particles.
This particularly implies that DM production during the thermalization of high-energy SM particles is inevitable if the DM is coupled to the SM sector.
All particles, including the DM,
must be produced through the cascades.
The DM production rate during the thermalization of SM particles can be estimated using the distribution of SM particles determined in this paper,
and the DM production cross section from the collision of SM particles.
This is demonstrated in Sec.~\ref{sec:application} by using a toy model.

\begin{table*}
\begin{center}
\begin{tabular}{c|ccccccccccc}
  & $e_f$ & $L_f$ & $u_{f'}$ & $u_3$ & $d_f$ & $Q_{f'}$ & $Q_3$  & $\phi$ & $g$ & $W$ & $B$  \\[.2em]
	\hline
	$100 \, R_s^{\rm (asym)}$ & 1.16  & 1.22  & 3.62  & 3.63  & 3.61  & 8.23  & 8.20
	 & 0.803  & 39.6  & 5.20  & 0.870
\end{tabular}
\end{center}
\caption{Asymptotic value of the fraction of distributions at a small momentum $R_s^{(\rm asym)}$.
}
\label{tab:1}
\end{table*}

%%%%%%%%%%%%%
\section{Numerical results}
\label{sec:numerical}
%%%%%%%%%%%%%

In this section, we first explain how to numerically determine the stationary solution to the Boltzmann equation. We then show our numerical results, which are consistent with the analytic result discussed in Sec.~\ref{sec:analytic}.

\subsection{Numerical methods}

We determine the stationary solution to the Boltzmann equation (\textit{i.e.}, $\partial f_s / \partial t = 0$) with a given source term for $s = \si$.
We again note that the Boltzmann equations are linear in terms of the distributions.
This allows us to take $\tilde{\Gamma} = 1$ without loss of generality by rescaling the normalization of distributions.
Moreover, we can take $\si$ independently for all species and
our results for $\si = \qty( e_f, L_f, u_{f'}, u_3, d_f, Q_{f'}, Q_3, \phi, g, W, B )$ cover all possible initial conditions for the source term.
We equally treat all generations except for the top quarks
and assume that high-energy particles are injected into all $f$ ($=1,2,3$) or $f'$ ($=1,2$) for each case of $\si$.
When we show our result for, \textit{e.g.}, $e_f$, we show that of one of the generations for $e_f$.

The stationary Boltzmann equation has two terms other than the source term: the terms proportional to $f_s(p)$ and those for the integral with a weight of $f_s (p')$ with $p' >p$.
This means that
the distribution at $p$ can be determined from the distribution for $p' > p$.
Therefore, we can determine a smaller momentum iteratively, starting from an appropriate boundary condition at the maximal momentum $p = p_0$.
The source term determines the boundary condition of the distributions at $p = p_0$, which we analytically determine in Sec.~\ref{sec:analytic} and Appendix~\ref{sec:appendixB}.

We discretize the momentum
to approximate the integral using a Simpson's rule method.
As we will see,
the particle distribution has a power-law dependence on $p$ for $p \ll p_0$ and on $p_0-p$ for $p \approx p_0$.
Therefore, it is convenient to discretize the momentum
in a logarithmic scale on $p$ for $p \ll p_0$ and on $p_0 - p$ for $p \approx p_0$.
Specifically, we use
\begin {align}
 &p_n = p_{\rm min} \exp \qty[
 \frac{\log (p_0/p_{\rm min})}{N_p} \sum_{i=1}^{i=n}
 \qty( 1-\tanh \qty( \frac{i-1/2 - N_p/2}{\sigma} ) )
 ],
 \\
 &\sigma = \frac{N_p}{\log \qty[ 2 \qty( p_0/T ) \log (p_0 / p_{\rm min}) /N_p ]},
\end {align}
for $n = 1, 2, \dots, N_p$, where $N_p$ is the number of grids.
One can check that $p_1 \simeq p_{\rm min} = 2T$ and $p_{N_p} \simeq p_0$ with exponentially small errors.
Here, $\sigma$ is chosen
such that the resolution of the momentum near $p \simeq p_0$ is approximately $T_0$, for example, $p_{N_p} - p_{N_p -1} \simeq T_0$.
As we will see shortly, there is no infrared divergence; hence, we do not require a small interval, particularly in the intermediate scale of $p$.
In our numerical calculation, we take $N_p = 10^4$ and $2p_0/p_{\rm min} = p_0 / T = 10^{12}$ unless otherwise stated.

The source term in \eqref{eq:boltzmann} is the delta function [see Eq.~\eqref{eq:source}]; therefore, we require an appropriate boundary condition or regularization in numerical calculations.
As discussed in Sec.~\ref{sec:analytic},
the stationary solution to the Boltzmann equation with the delta-function source term can be analytically calculated for $p \approx p_0$.
The distribution for a primary particle $\si$ is proportional to $\sqrt{p_0/(p_0 - p)}$ for $\si = L_f, u_{f'}, u_3, d_f, Q_{f'}, Q_3, \phi, g, W$.
In our numerical calculation,
we use this analytic form of primary particle distribution for $p_n \in (p_{\rm ini}, p_0)$ with $p_{\rm ini}$ determined via $(p_0 -p_{\rm ini})/p_0 \simeq 10^{-3}$.
One can also take analytic distributions for other (secondary) particles, which are derived in Appendix~\ref{sec:appendixB}. However, these are subdominant at $p \approx p_0$ and are mainly produced from the primary particle; hence, they can be omitted for $p > p_{\rm ini}$.
We find that the resulting distributions for $p \lesssim p_{\rm ini}$ do not change for a different choice in $p_{\rm ini}$ within a numerical uncertainty if $(p_0 -p_{\rm ini})/p_0 \ll 1$ is satisfied.
For the case of $\si = e_f$ and $B$,
the distribution of a primary particle $\si$ is proportional to a delta function. In this case,
we simply substitute the delta function into the Boltzmann equation.
Specifically, for the case of $\si = e_f$, we include
$(2\pi)^3/(p^2 \nu_B) \, 3 \gamma_{e_f \leftrightarrow B e_f} (p_0;p,p_0-p) C_{e_f}'$ in \eqref{eq:boltzmann-photon}
instead of imposing boundary conditions for $\si = e_f$.
Including these terms, we solve the Boltzmann equation from $p \approx p_0$ to a smaller $p$.
The case with $\si = B$ is similar to the case with $\si = e_f$. In this case, we should include
$(2\pi)^3/(p^2 \nu_s) \, 2 d_s^{(2)}d_s^{(3)} \gamma_{B \leftrightarrow s \bar{s}} (p_0;p,p_0-p) C_{B}'$ in \eqref{eq:boltzmannparticle} (or correspondingly, Eqs.~(\ref{eq:boltzmannuf}), (\ref{eq:boltzmannQf}), (\ref{eq:boltzmannef}), (\ref{eq:boltzmannLf}), and (\ref{eq:f_phi})).

For a consistency check,
we utilize a different algorithm that simply uses a constant initial condition (fixed boundary condition) at $p = p_0$ for a particle species $\si$.
This method is used in Refs.~\cite{Drees:2021lbm,Drees:2022vvn}.
We find that
the results of this method are consistent with those determined by the above methods within a numerical uncertainty.

The splitting rate $\gamma_{s\leftrightarrow s' s''}(p;k, p-k)$ is proportional to $k^{-3/2}$ for soft gauge boson emission with a small momentum $k$. This implies an infrared divergence for the integral over $k$.
However, this is automatically regulated when we include all terms in the Boltzmann equation
because the emission of soft gauge bosons with energy $\epsilon p$ from energy $p$ is cancelled by the contribution from
the emission of soft gauge bosons with energy $p+\epsilon p$ in the limit of small $\epsilon$. Because we expect that the distribution function is continuous,
these contributions cancel each other out,
and the splitting terms in the Boltzmann equation do not exhibit infrared divergence.
Therefore, we do not (in principle) need to include an infrared cutoff to perform the numerical calculations.
Still, we must take particular care with the integral in a small $k \ll p$.
Let us consider the gluon self-splitting process as an example. The (apparent) IR divergent parts are given by
\begin{align}
  \int_0^{\epsilon p}  \dd k \,
    \gamma_{g \leftrightarrow gg} \bigl(p; k, p-k \bigr) \,
    f_g(p)
    -
  \int_0^{\epsilon p} \dd k \,
    \gamma_{g \leftrightarrow gg} \bigl(p+k; p, k \bigr) \,
    f_g(p+k)
\end{align}
where we focus on $k \in (0, \epsilon p)$ with $\epsilon \ll 1$.
Because $\epsilon \ll 1$,
we can approximate $f_g(p+k) \simeq f_g(p) + k f_g'(p)$.
Then, we obtain
\begin{align}
  f_g (p) \int_0^{\epsilon p} \dd k \,
    \qty[ \gamma_{g \leftrightarrow gg} \bigl(p; k, p-k \bigr) - \gamma_{g \leftrightarrow gg} \bigl(p+k; p, k \bigr) ]
    -
  f_g'(p) \int_0^{\epsilon p} \dd k \, k \,
    \gamma_{g \leftrightarrow gg} \bigl(p+k; p, k \bigr).
\end{align}
These integrals are regular and can be performed analytically or numerically.
In our discrete momentum method,
we must determine $f_g'(p_i)$ from the information of $f_g(p_j)$ with $j > i$.
One may simply calculate the derivative from $f_g(p_{j})$ with $j > i$, and interpolate it to $p = p_i$.
However, this may result in artificial instability for the numerical calculation.
This is because an error on the derivative at $p_j = p_{i+1}$ leads to an error on $f_g(p_i)$, which results in a larger error on the derivative at $p_i$.
This enhances the error on $f_g(p)$ at a small $p$ when we iteratively determine $f_g(p)$ from a large $p$.
To avoid this artificial instability,
we calculate the derivatives for $f_g(p_{j})$ with $j = i+1, i+2, \dots, i+100$ and take their averaged value to approximate $f_g'(p_i)$.
In other words, we discretize the derivatives of the distributions by $N_p/100 = 100$ grids rather than $N_p$ grids. This is still accurate for our purpose because the derivatives of the distributions do not change drastically within $p \in (p_i, p_{i+100})$.

We emphasize that this procedure for the (apparent) IR divergence
allows us to take a larger momentum grid than the physical IR cutoff of $\mathcal{O}(T)$.
This in turn allows us to start from a significantly larger momentum $p_0$ than $T$.
In fact, we can take $p_0 / T = 10^{12}$ or larger,
even for a relatively small number of grids, $N_p = 10^4$.

\subsection{Numerical results and scaling solution}
\label{sec:numericalscaling}

%%%%%%%%%%%%%%%%%%%%%%%%%%%%%%%%%%%%%%%%%%%%%%%%%%%%%%%%%%%%%%%%%%%%%%%%%%%%%%%%%%%%%%%%%%%%%%%%%%%%

\begin{figure}[t]
	\centering
 	\includegraphics[width=0.4\linewidth]{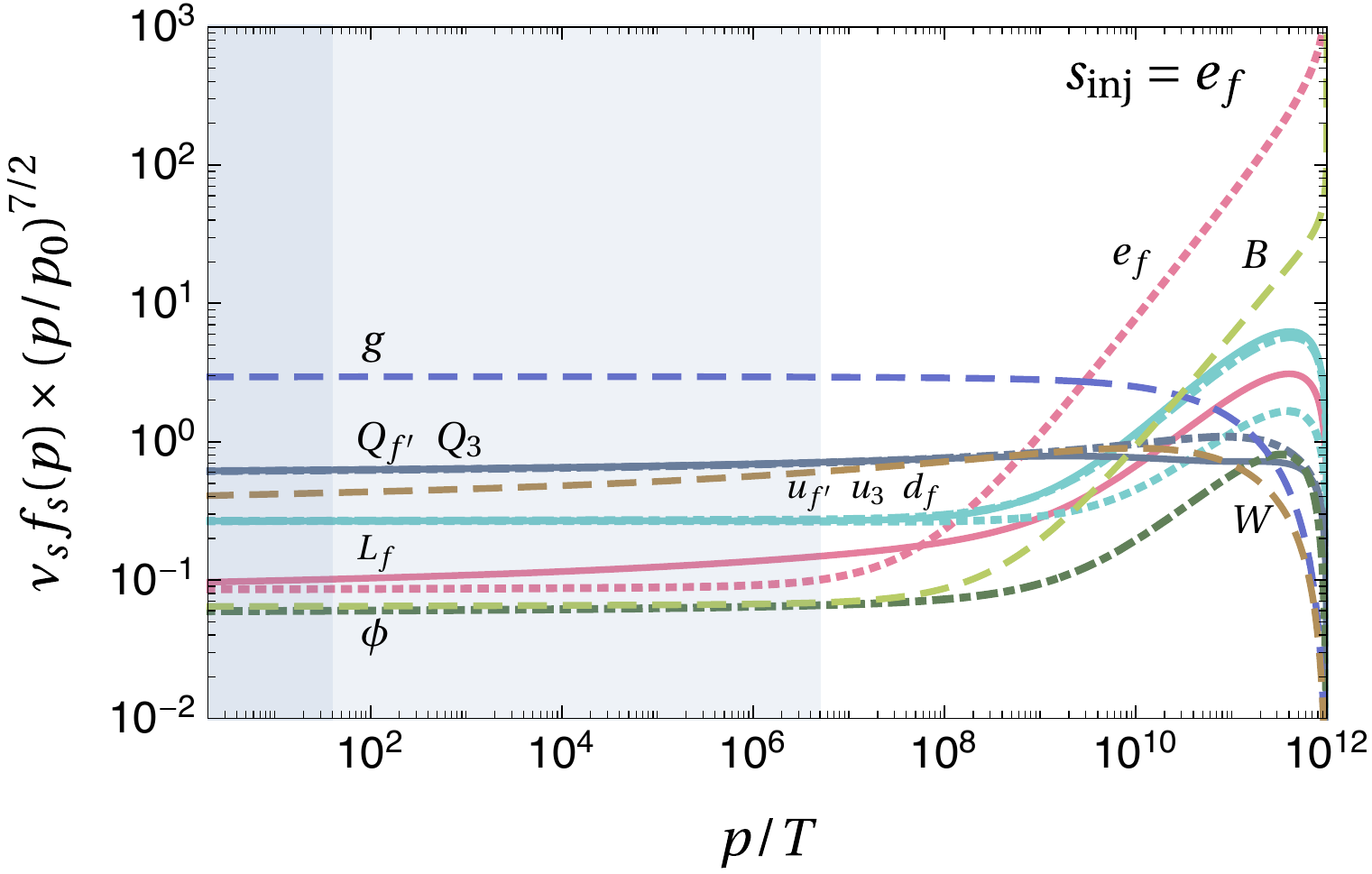}
 	\hspace{0.4cm}
 	\includegraphics[width=0.4\linewidth]{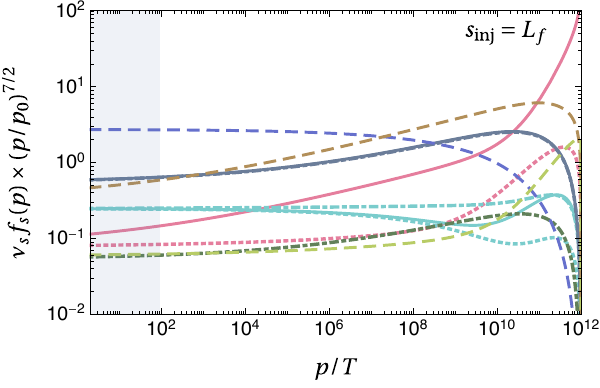}
 	\vspace{0.2cm}\\
 	\includegraphics[width=0.4\linewidth]{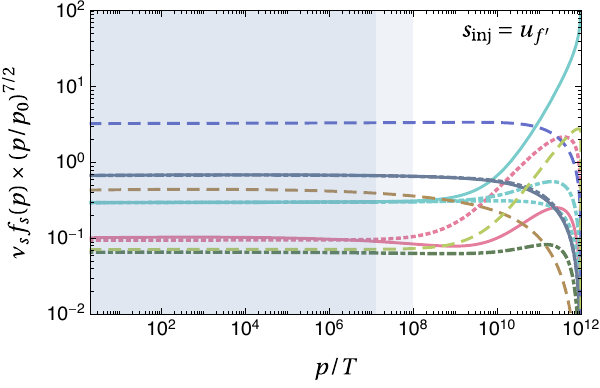}
 	\hspace{0.4cm}
 	\includegraphics[width=0.4\linewidth]{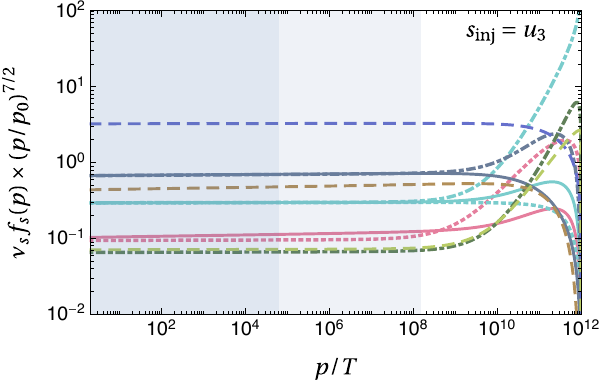}
 	\vspace{0.2cm}\\
 	\includegraphics[width=0.4\linewidth]{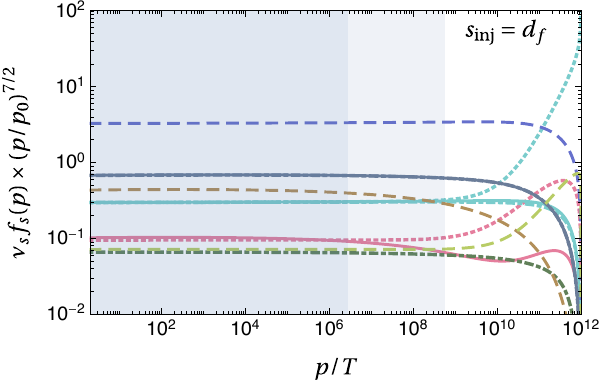}
 	\hspace{0.4cm}
 	\includegraphics[width=0.4\linewidth]{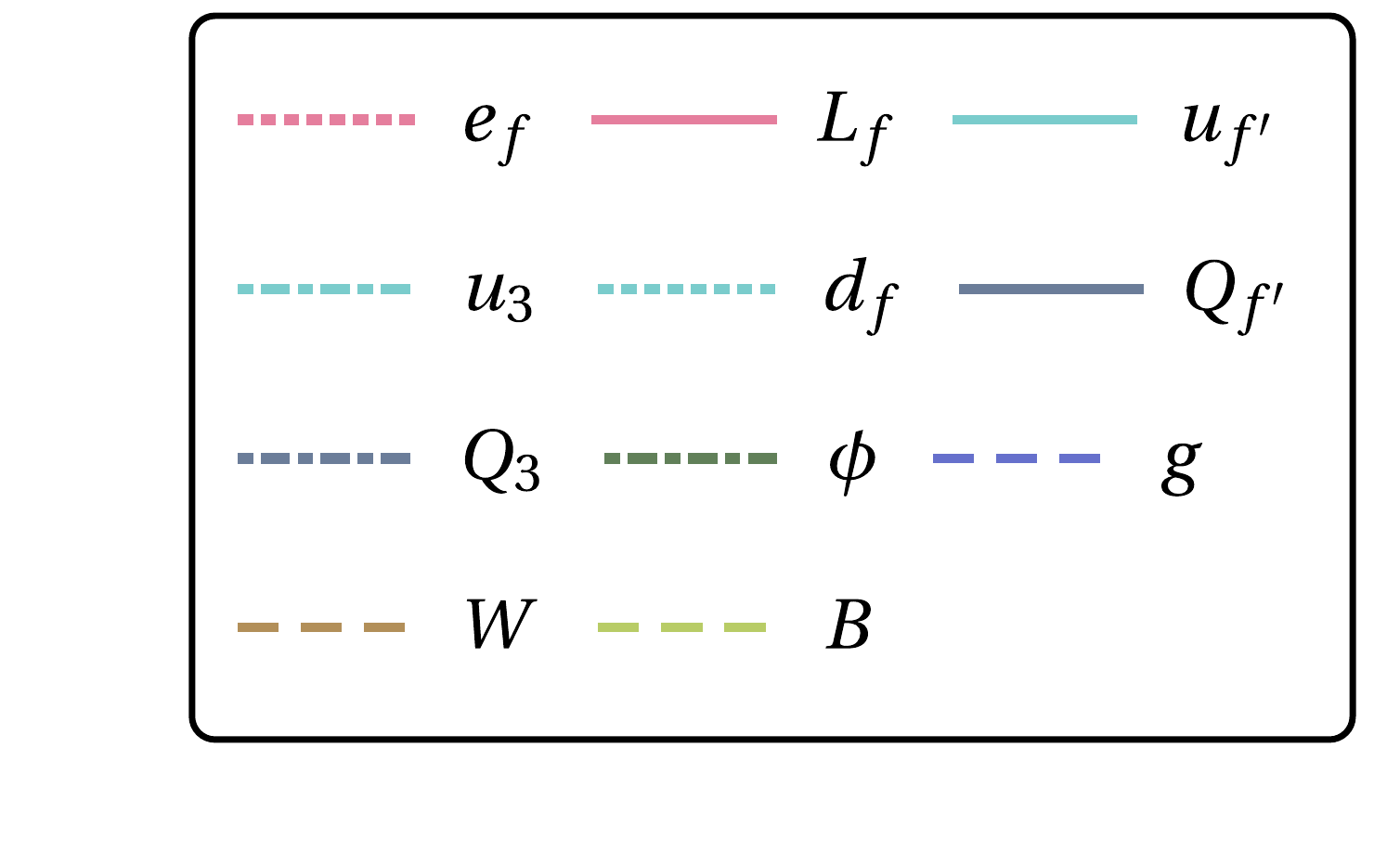}
	\caption{
	Particle distributions for each species $s$, rescaled by $(p/p_0)^{7/2}$. The legends for $s$ are shown in the lower right figure.
	The primary particle $\si$ is injected solely for $e_f$ (upper left), $L_f$ (upper right),
	$u_{f'}$ (middle left), $u_3$ (middle right),
	and $d_f$ (lower left).
	The dark and light shaded regions represent $p < p_{\rm asym}^{(s)}$ and $p < p_{\rm asym'}^{(s)}$, respectively.
    }
	\label{fig:f-1}
\end{figure}

\begin{figure}[t]
	\centering
 	\includegraphics[width=0.4\linewidth]{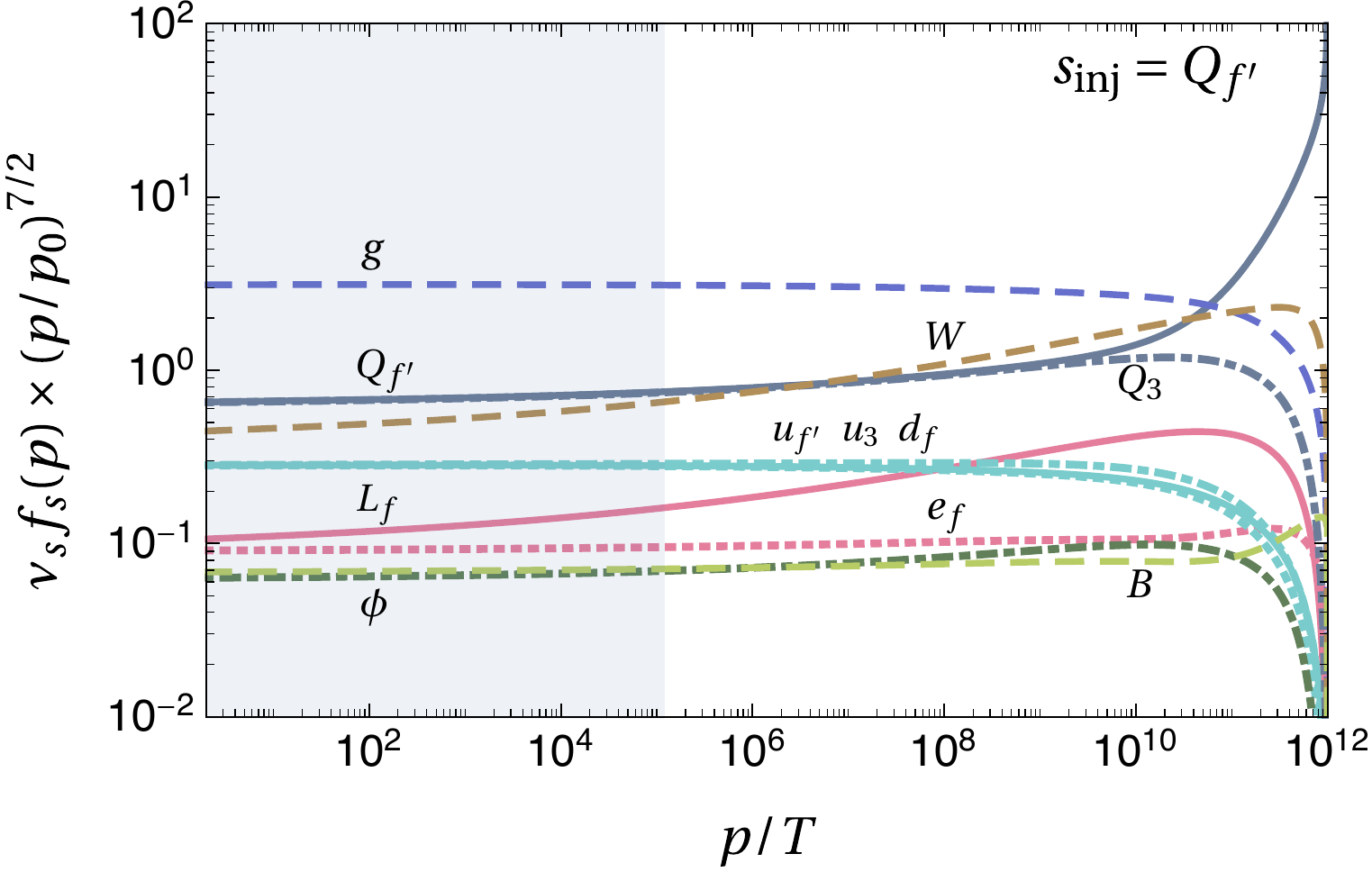}
 	\hspace{0.4cm}
 	\includegraphics[width=0.4\linewidth]{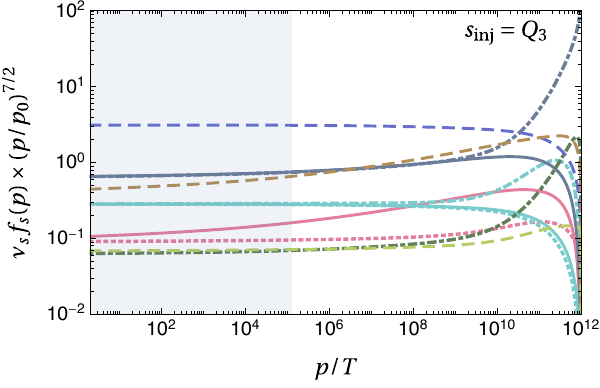}
 	\vspace{0.3cm}\\
 	\includegraphics[width=0.4\linewidth]{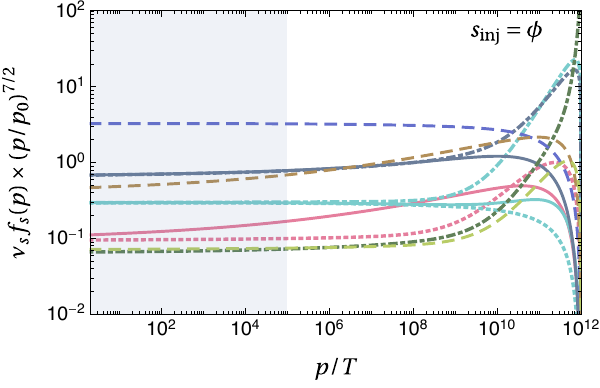}
 	\hspace{0.4cm}
 	\includegraphics[width=0.4\linewidth]{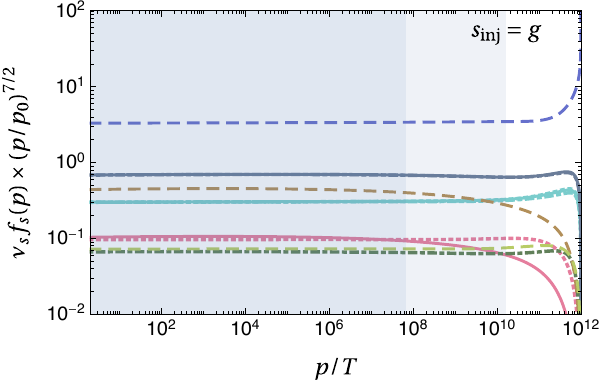}
 	\vspace{0.3cm}\\
 	\includegraphics[width=0.4\linewidth]{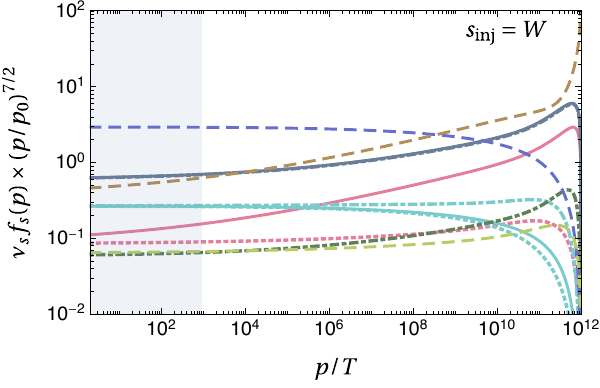}
 	\hspace{0.4cm}
 	\includegraphics[width=0.4\linewidth]{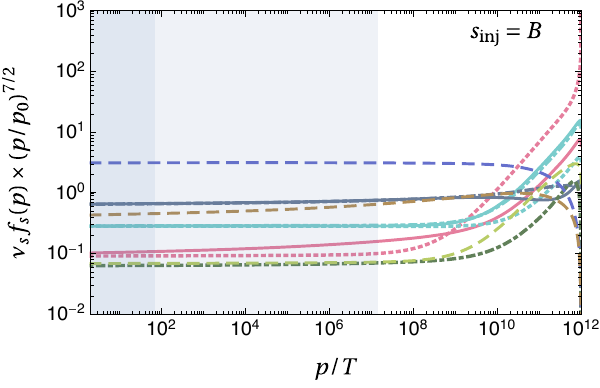}
	\caption{
	Same as Fig.~\ref{fig:f-1}, but the primary particle is injected solely for $Q_{f'}$ (upper left), $Q_3$ (upper right),
	$\phi$ (middle left), $g$ (middle right),
	$W$ (lower left), and $B$ (lower right).
    }
	\label{fig:f-2}
\end{figure}

%%%%%%%%%%%%%%%%%%%%%%%%%%%%%%%%%%%%%%%%%%%%%%%%%%%%%%%%%%%%%%%%%%%%%%%%%%%%%%%%%%%%%%%%%%%%%%%%%%%%

The numerical results for the particle distributions weighted by $(p/p_0)^{7/2}$ are shown in Fig.~\ref{fig:f-1} for $\si = e_f$ (top left), $L_f$ (top right), $u_{f'}$ (middle left), $u_3$ (middle right), and $d_f$ (bottom left)
and in Fig.~\ref{fig:f-2} for $\si = Q_{f'}$ (top left), $Q_3$ (top right),
$\phi$ (middle left), $g$ (middle right), $W$ (bottom left), and $B$ (bottom right).
As stated, we use $T = 10^3 \GeV$, $p_0/ T = 10^{12}$, and $N_p = 10^4$,
where the dependence of our results on $T$ (or equivalently, $p_0$) is only logarithmic through the running of the gauge couplings.
We take $\tilde{\Gamma} = 1$ without loss of generality, where our results for $f_s (p)$ should be rescaled by $\tilde{\Gamma}$ for other values of $\tilde{\Gamma}$.
In all cases except $\si = B$,
the injected particle is dominant at $p \approx p_0$,
and other secondary particles are produced with the suppression factors discussed in Appendix~\ref{sec:appendixB} (see Fig.~\ref{fig:flow1}).
Our numerical results confirm that
all particle distribution scale with $f \propto p^{-7/2}$ for $p \ll p_0$ for any injected particles.
For a smaller $p$, the particle distribution reaches a scaling solution, which is independent of the initial injected particle.

\begin{figure}[t]
	\centering
 	\includegraphics[width=0.4\linewidth]{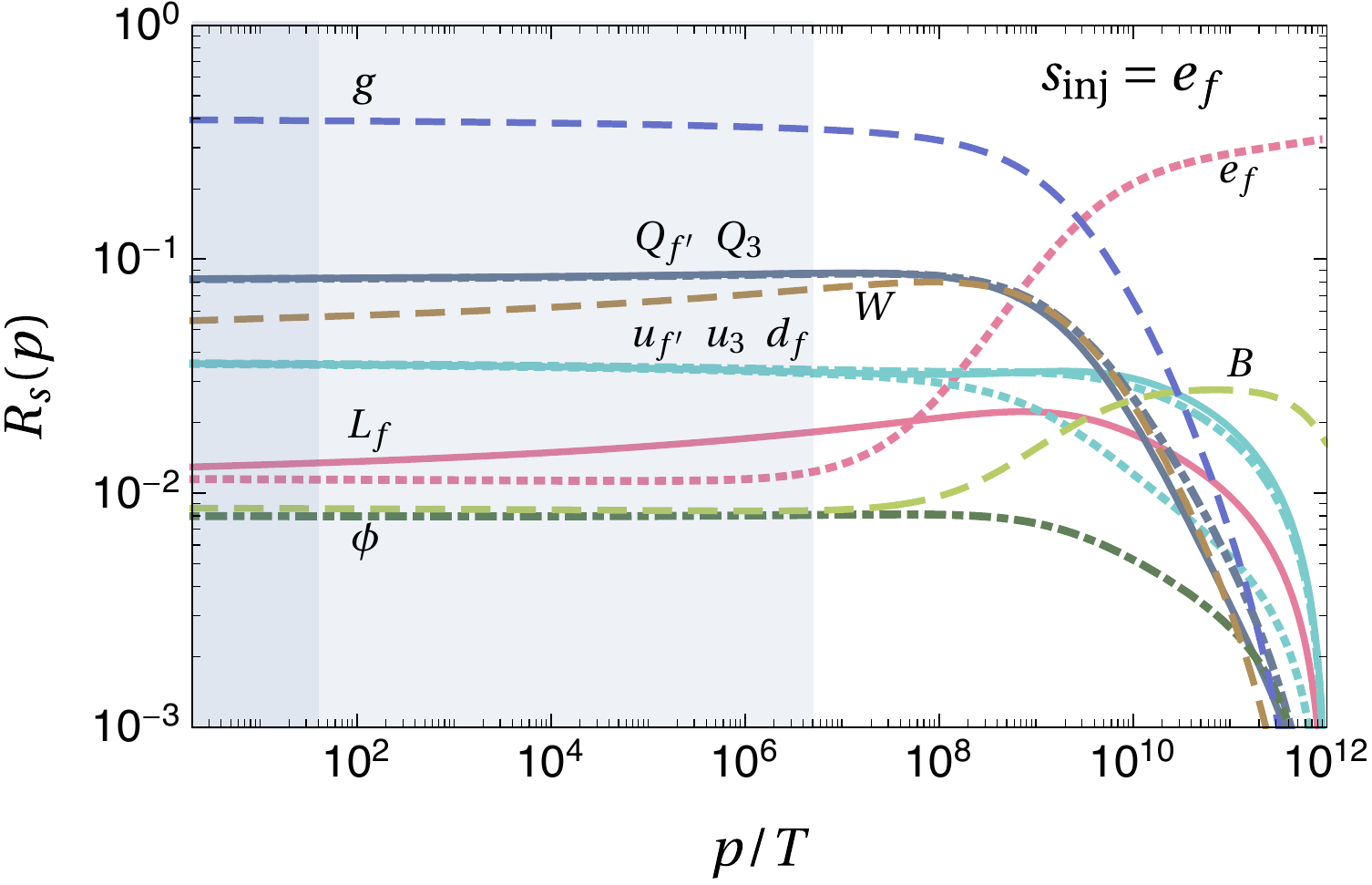}
 	\hspace{0.4cm}
 	\includegraphics[width=0.4\linewidth]{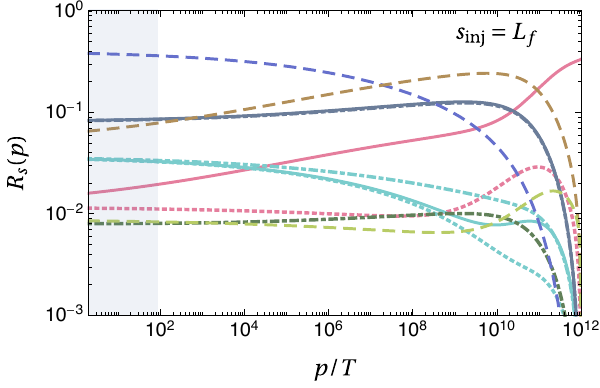}
 	\vspace{0.2cm}\\
 	\includegraphics[width=0.4\linewidth]{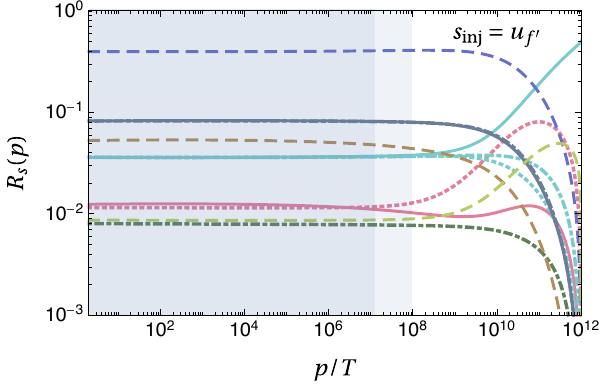}
 	\hspace{0.4cm}
 	\includegraphics[width=0.4\linewidth]{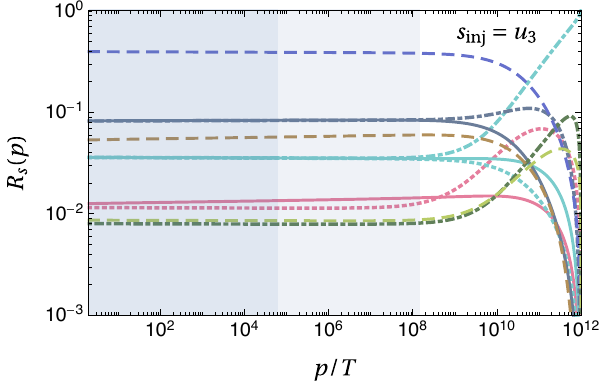}
 	\vspace{0.2cm}\\
 	\includegraphics[width=0.4\linewidth]{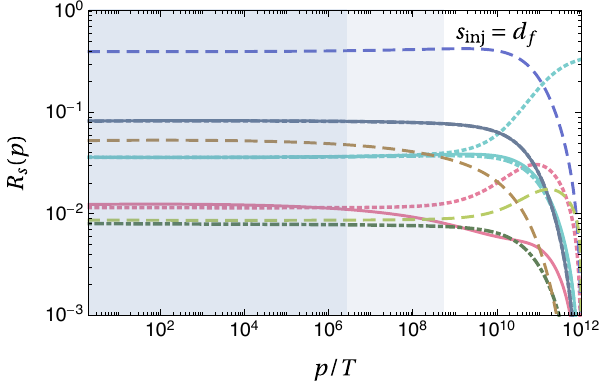}
 	\hspace{0.4cm}
 	\includegraphics[width=0.4\linewidth]{legends}
	\caption{
	Fraction of species $s$ at an energy $p$, $R_s(p)$. The primary particle is injected solely for $e_f$ (upper left), $L_f$ (upper right),
	$u_{f'}$ (middle left), $u_3$ (middle right),
	and $d_f$ (lower left).
	The dark and light shaded regions represent $p < p_{\rm asym}^{(s)}$ and $p < p_{\rm asym'}^{(s)}$, respectively.
    }
	\label{fig:R-1}
\end{figure}

\begin{figure}[t]
	\centering
 	\includegraphics[width=0.4\linewidth]{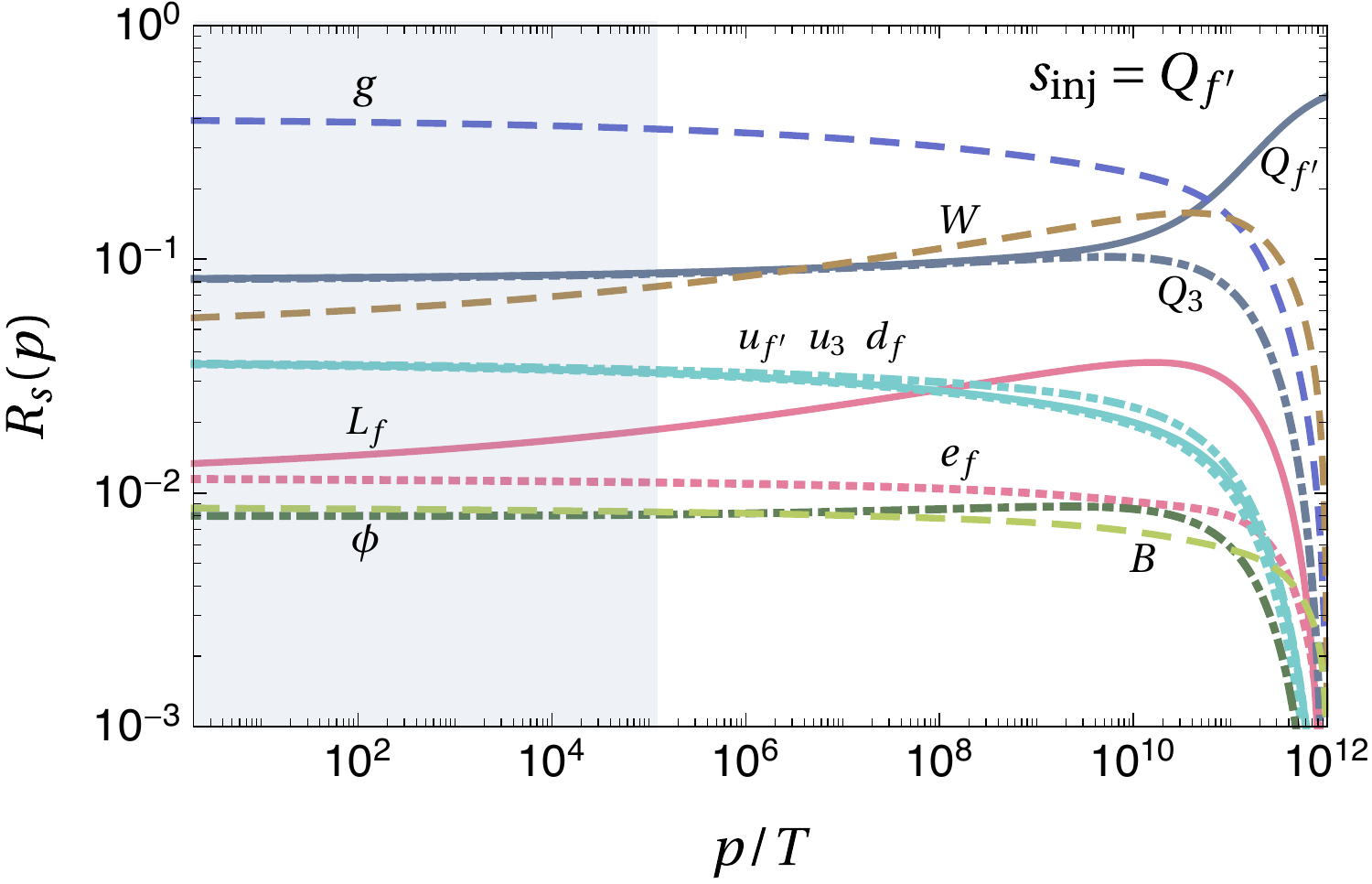}
 	\hspace{0.4cm}
 	\includegraphics[width=0.4\linewidth]{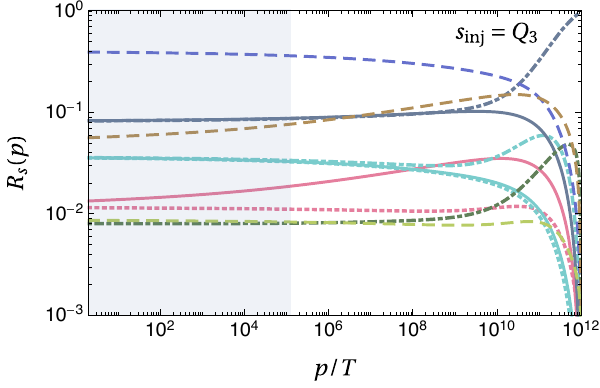}
 	\vspace{0.4cm}\\
 	\includegraphics[width=0.4\linewidth]{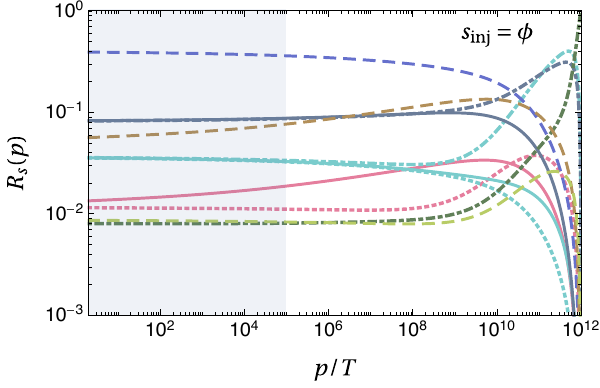}
 	\hspace{0.4cm}
 	\includegraphics[width=0.4\linewidth]{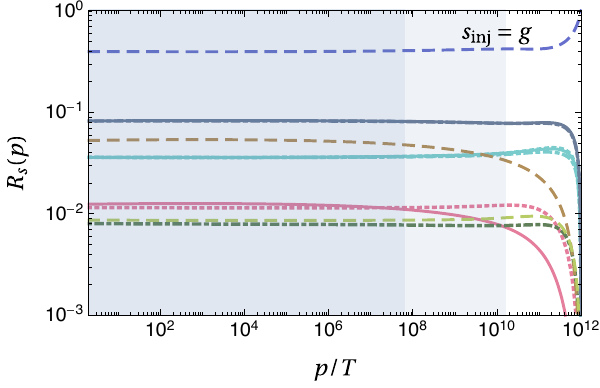}
 	\vspace{0.4cm}\\
 	\includegraphics[width=0.4\linewidth]{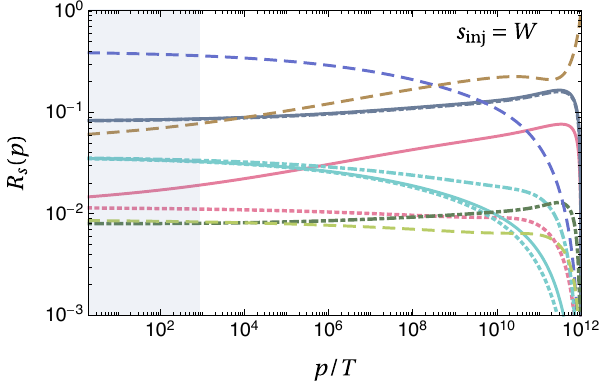}
 	\hspace{0.4cm}
 	\includegraphics[width=0.4\linewidth]{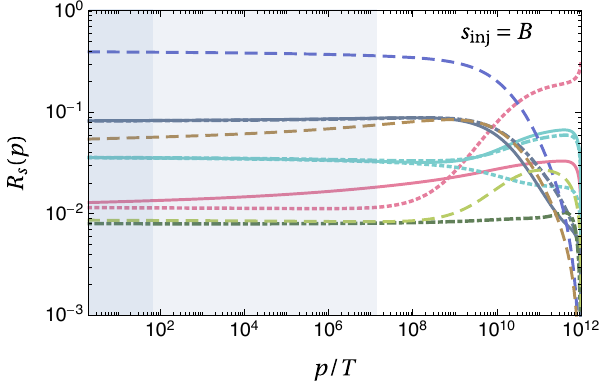}
	\caption{
	Same as Fig.~\ref{fig:R-1}, but the primary particle is injected solely for $Q_{f'}$ (upper left), $Q_3$ (upper right),
	$\phi$ (middle left), $g$ (middle right),
	$W$ (lower left), and $B$ (lower right).
    }
	\label{fig:R-2}
\end{figure}

To determine the dominant particle species during the splitting process, we plot $R_s(p)$ given by \eqref{eq:Rs} as a function of $p$ in Figs.~\ref{fig:R-1} and \ref{fig:R-2} for each $\si$.
We can see that $R_s$ is asymptotic to a certain value $R_s^{\rm (asym)}$ for a sufficiently small $p/p_0$ for any $\si$.
The asymptotic values are consistent with the analytic result derived in Sec.~\ref{sec:analytic},
as shown in Tab.~\ref{tab:1}.
In particular, the gluon dominates the number and energy of SM particles in the scaling regime for any $\si$.

%%%%%%%%%%%%%%%%%%%%%%%%%%%%%%%%%%%%%%%%%%%%%%%%%%%%%%%%%%%%%%%%%%%%%%%%%%%%%%%%%%%%%%%%%%%%%%%%%%%%

%%%%%%%%%%%%%%%%%%%%%%%%%%%%%%%%%%%%%%%%%%%%%%%%%%%%%%%%%%%%%%%%%%%%%%%%%%%%%%%%%%%%%%%%%%%%%%%%%%%%

The energy threshold for the scaling solution is represented by $p_{\rm asym}^{(\si)}$,
where
$R_s(p) \simeq R_s^{\rm (asym)}$ for all $s$ for $p < p_{\rm asym}^{(\si)}$.
Namely, $p_{\rm asym}^{(\si)}$ represents
the energy below which
$R_s(p)$ is equal to the scaling solution within $10\%$ for any $s$
for the case in which a species $\si$ is injected at $p = p_0$.
The particles charged under SU(2) tend to require a relatively small $p_{\rm asym}^{(\si)}$
and cannot reach the scaling solution in the case of $p_0 / T = 10^{12}$.
For those cases, we also calculate the stationary solution to the Boltzmann equation with $p_0 / T = 10^{14}$.
The cases with $\si = L_f$ and $W$ do not reach the scaling solution even for $p_0 / T = 10^{14}$, whereas the cases with $\si = Q_{f'}$, $Q_3$, and $\phi$, reach the scaling solution at $p_{\rm asym}^{(\si)} = \mathcal{O}(10^{-13}) p_0$.

We also define $p_{\rm asym'}^{(\si)}$ such
that
$R_s(p) \simeq R_s^{\rm (asym)}$ for $p < p_{\rm asym'}^{(\si)}$ for all $s$ except $s = L_f, W$.
This represents the scale below which the dominant components, such as gluons, reach the scaling regime
while particles with only the slowest SU(2) interactions may not.
The results are summarized in Tabs.~\ref{tab:3} and \ref{tab:4}.
In Fig.~\ref{fig:f-1}, \ref{fig:f-2}, \ref{fig:R-1}, and \ref{fig:R-2},
the dark and light shaded regions  represent $p < p_{\rm asym}^{(s)}$ and $p < p_{\rm asym'}^{(s)}$, respectively.
The difference between $p_{\rm asym}^{(\si)}$ and $p_{\rm asym'}^{(\si)}$
represents how slowly $s = L_f$ and $W$ reach the scaling solution, even after the other species achieve this.

\begin{table*}
\begin{center}
\begin{tabular}{c|ccccccc}
  & $e_f$ & $L_f$ & $u_{f'}$ & $u_3$ & $d_f$ & $Q_{f'}$ & $Q_3$
	\\[.2em]
	\hline
	$\tilde{f}_{\rm tot, \si}^{\rm (asym)}$ & $7.46$  & $7.56$  & $8.23$  & $8.18$  & $8.25$  & $7.89$  & $7.88$
	\\
	$p_{\rm asym}^{(\si)}/p_0$ & $4 \times 10^{-11}$  & $< 10^{-14}$  & $1 \times 10^{-5}$  & $6 \times 10^{-8}$  & $3 \times 10^{-6}$ & $4 \times 10^{-13}$  & $4 \times 10^{-13}$
	\\
	$p_{\rm asym'}^{(\si)}/p_0$ & $5 \times 10^{-6}$  & $9 \times 10^{-11}$  & $1 \times 10^{-4}$  & $2 \times 10^{-4}$  & $6 \times 10^{-4}$ & $1 \times 10^{-7}$  & $1 \times 10^{-7}$
\end{tabular}
\end{center}
\caption{Asymptotic value of total distributions and the typical energy scales of the scaling solution $p_{\rm asym}^{(\si)}$ and $p_{\rm asym'}^{(\si)}$ for $\si = (e_f, L_f, u_{f'}, u_3, d_f, Q_{f'}, Q_3)$.
}
\label{tab:3}
\end{table*}

\begin{table*}
\begin{center}
\begin{tabular}{c|cccc}
  & $\phi$ & $g$ & $W$ & $B$  \\[.2em]
	\hline
	$\tilde{f}_{\rm tot, \si}^{\rm (asym)}$ & $8.67$  & $8.33$  & $8.04$  & $7.92$
	\\
	$p_{\rm asym}^{(\si)}/p_0$  & $3 \times 10^{-13}$ & $7 \times 10^{-5}$  & $<  10^{-14}$  & $7 \times 10^{-11}$
	\\
	$p_{\rm asym'}^{(\si)}/p_0$  & $1 \times 10^{-7}$ & $2 \times 10^{-2}$  & $9 \times 10^{-10}$  & $1 \times 10^{-5}$%	\\
\end{tabular}
\end{center}
\caption{Same as Tab.~\ref{tab:1} but for $\si =( \phi, g, W, B)$.
}
\label{tab:4}
\end{table*}

In Sec.~\ref{sec:analyticsmall}, we discuss that the asympotic value of total distribution $\tilde{f}_{\rm tot}^{\rm (asym)}$ is independent of $\si$.
To confirm this,
we define
an asymptotic value of total distributions for each $\si$ such as
\begin{align}
 f_{\rm tot, \si}^{\rm (asym)} (p)
 &\equiv
 \left. \sum_s \nu_s f_s(p)
 \right\vert_{p < p_{\rm asym}^{(s_{\rm inj})}},
 \label{eq:totaldist2}
 \\
 &\equiv
 \tilde{\Gamma} \tilde{f}_{\rm tot, \si}^{\rm (asym)}
 \qty( \frac{p}{p_0} )^{-7/2}.
 \label{eq:tildetotal2}
\end{align}
The proportionality constant is determined from numerical calculations for each $\si$ and is shown in Tabs.~\ref{tab:3} and \ref{tab:4} for $\tilde{\Gamma} = 1$.
We also show the results of $\si = L_f$ and $W$
even though they do not reach the scaling solution.
For all cases, $\tilde{f}_{\rm tot, \si}^{\rm (asym)}$ is the same with each other within an error of order $10 \%$.
The averaged value is about $8.04$,
which is consistent with the rough estimation of \eqref{eq:tildeftotest}.%
\footnote{
For the case with only gluons, namely for the pure SU(3) gauge theory, we obtain $\tilde{f}_{\rm tot}^{\rm (asym)} \simeq 4.08$
from our numerical calculation. This is smaller than that for the full SM case by a factor of about two. The difference comes from the fact that some fraction of energy is transferred into the other SM particles, which have slower thermalization processes. The thermalization rate of the whole sector is therefore smaller in the full SM case and the resulting asymptotic value $\tilde{f}_{\rm tot}^{\rm (asym)}$ becomes larger than the case only with gluons.
}
The derivation for different $\si$ may come from numerical errors and the fact that the scaling solution is not an exact for a finite $p/p_0$.
In particular,
a relatively large deviation for $\si = L_f$ come from the fact that the system does not reach the scaling solution even for $p = 10^{-14} p_0$.
For example, the asymptotic values $\tilde{f}_{\rm tot, \si}^{\rm (asym)}$
for $\si = (L_f, W)$ change from ($7.13, \, 7.60$) for $p/p_0 = 10^{-12}$
to ($7.57, \, 8.04$) for $p/p_0 = 10^{-14}$.

\begin{figure}[t]
	\centering
 	\includegraphics[width=0.4\linewidth]{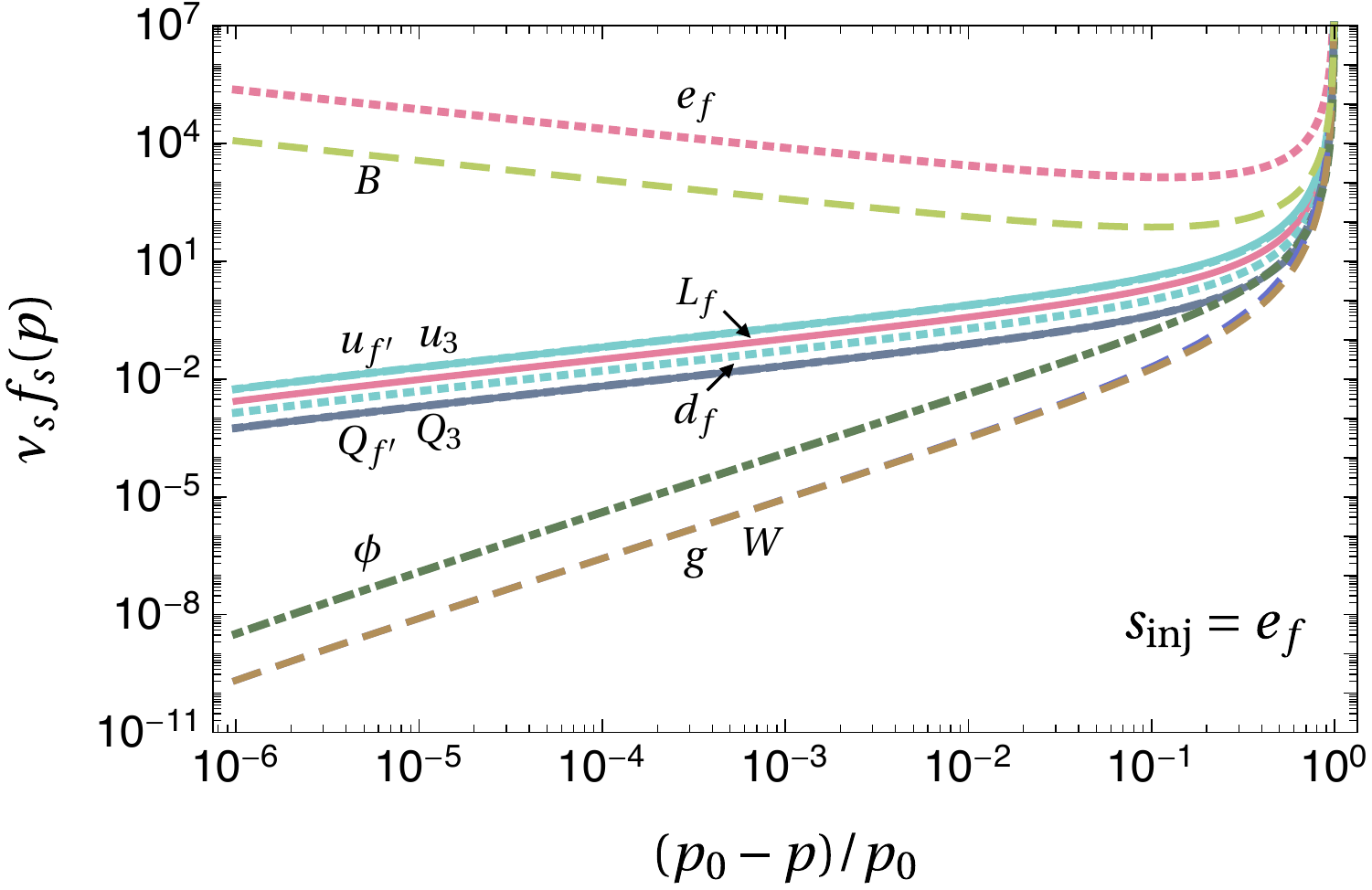}
 	\hspace{0.4cm}
 	\includegraphics[width=0.4\linewidth]{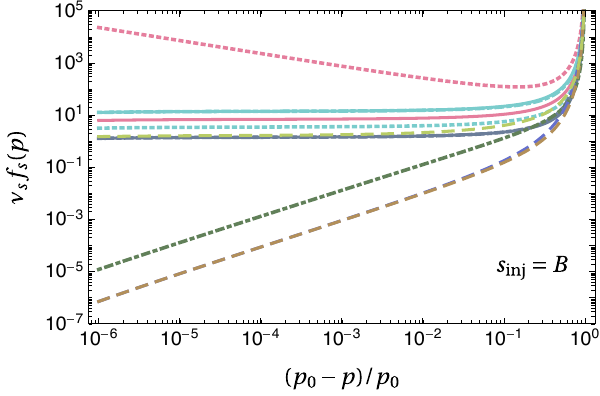}
	\caption{
	Particle distributions for $p \approx p_0$. The primary particle is injected solely for $e_f$ (left) and $B$ (right). The asymptotic behaviors for $p \approx p_0$ are consistent with the analytic result shown in Fig.~\ref{fig:flow1}.
    }
	\label{fig:p0}
\end{figure}

Although the asymptotic behavior for $p \ll p_0$ is similar in all cases,
the initial cascading process at $p \approx p_0$ is different.
These features are, however, consistent with the analytic calculations derived in Sec.~\ref{sec:analytic} and Appendix~\ref{sec:appendixB}.
We show a couple of examples of distributions at $p \approx p_0$ in Fig.~\ref{fig:p0},
where $\si = e_f$ (left panel) and $B$ (right panel).
In the figure, we do not show the delta-function distribution for $e_f$ and $B$.
The distributions of all particles have a consistent dependence with Fig.~\ref{fig:flow1}.
The cascading process at $p \approx p_0$ can also be seen in Figs.~\ref{fig:f-1} and \ref{fig:f-2}, which are consistent with Fig.~\ref{fig:flow1}.

%%%%%%%%%%%%%
\section{Application to non-thermal DM production}
\label{sec:application}
%%%%%%%%%%%%%

In this section, we briefly explain how our results can be used to calculate the DM abundance from non-thermal production during the thermalization process.
We can improve the qualitative estimation of Ref.~\cite{Harigaya:2014waa} to a quantitative level.

Suppose that DM $\chi$ is produced from
the reaction of $s_1 s_2 \to \chi \chi$.
The Boltzmann equation is given by
\begin{align}
 \frac{\dd}{\dd t} n_\chi + 3 H n_\chi
&=
  2 \int \dd {\rm Lips}
  \abs{\mathcal{M}}^2 \nu_{s_1} f_{s_1} (p_1)
  \nu_{s_2} f_{s_2} (p_2),
  \\
 &= 2 \int \frac{\dd^3 p_1}{(2 \pi)^3} \frac{\dd^3 p_2}{(2\pi)^3}
 \nu_{s_1} f_{s_1} (p_1)
  \nu_{s_2} f_{s_2} (p_2)
  \sigma v,
\end{align}
where $\mathcal{M}$ is the amplitude of the scattering process $s_1 s_2 \to \chi \chi$ and
\begin{align}
 &\dd {\rm Lips} \equiv
  \dd\Pi_1 \dd \Pi_2 \dd \Pi_3 \dd \Pi_4 (2 \pi)^4 \delta(p_1 + p_2 - p_3 - p_4),
  \\
 &\dd \Pi_i \equiv \frac{\dd^3 p_i}{(2\pi)^3 2 E_i}.
\end{align}
The distribution $f_s$ is given by a combination of
the thermal and non-thermal parts, such as
\begin{align}
 &\nu_s f_s(p) = \nu_s f_s^{\rm (th)}(p) + \nu_s f_s^{\rm (Non-th)}(p),
 \\
 &f_s^{\rm (th)} (p) = \frac{1}{e^{p/T} \pm 1}.
\end{align}
Here, the non-thermal part is given by
\begin{align}
 \nu_s f_s^{\rm (Non-th)} (p) &= R_s (p) f_{\rm tot} (p)
 \\
 &\simeq
 \frac{ 2 \pi^2 \rho_I(t) \Gamma_I }{p_0^{7/2} T^{3/2}(t)} R_s^{\rm (asym)} \tilde{f}_{\rm tot}^{\rm (asym)} \qty( \frac{p}{p_0} )^{-7/2},
\end{align}
for $p \gtrsim T$,
where we use Eq.~(\ref{eq:tildetotal}) and $\tilde{\Gamma} = 2 \pi^2 \rho_I \Gamma_I / (p_0^{7/2} T^{3/2})$
and assume $p < p_{\rm asym}^{(\si)}$
in the final line such that
the production process is dominated by scattering with energy $p$ in the scaling regime.
In this form, the time dependence originates from $\rho_I$ and $T$.
Note that $\tilde{f}_{\rm tot}^{\rm (asym)}$ only depends on $T(t)$ logarithmically; hence, we can neglect its dependence on the time variable.
We can solve the Boltzmann equation for $\chi$ to obtain its abundance by
substituting $R_s^{\rm (asym)}$ and $\tilde{f}_{\rm tot}^{\rm (asym)}$
($\simeq 21\,\text{-}\,23$) from Tabs.~\ref{tab:3} and \ref{tab:4}.

At the end of reheating, we have $T = T_{\rm RH} \simeq (90/(g_* \pi^2))^{1/4} \sqrt{\Gamma_I \Mpl}$, where $g_*$ is the effective number of relativistic degrees of freedom.
If the primary particles are produced from inflaton decay, we have $\rho_I = 3 H^2 \Mpl^2 \simeq 3 \Gamma_I^2 \Mpl^2$.

Now, we shall calculate the DM abundance in a toy model.
We assume that the cross section for DM production is given by $\sigma = \alpha_{\chi}^2 /s$ for $s > 4 m_\chi^2$, where $\alpha_{\chi}$ is a coupling constant, and $s$ is the center-of-mass energy.
When $m_\chi^2 / T < p_0$,%
\footnote{
When $m_\chi^2 / T > p_0$,
DM can be produced mainly by scattering among the high-energy particles from the cascade~\cite{Harigaya:2014waa}.
}
DM can be produced mainly by the scattering between the thermal plasma and high-energy particles from the cascade, such as
\begin{align}
 \frac{\dd}{\dd t} n_\chi + 3 H n_\chi
&=
\frac{ 2 \pi^2 \rho_I(t) \Gamma_I }{p_0^{7/2} T^{3/2}(t)}
\qty( \nu_{s_1} R_{s_2}^{\rm (asym)} +
\nu_{s_2} R_{s_1}^{\rm (asym)} )
 \tilde{f}_{\rm tot}^{\rm (asym)}
  \nonumber\\
  &\qquad \qquad \qquad \times
 \frac{1 }{4\pi^4}
  \int_0^\infty p_1^2 \dd p_1
 \int_T^{p_0} p_2^2 \dd p_2 \int_{-1}^1 \dd \cos \theta \,
  e^{-p_1/T} \qty( \frac{p_2}{p_0} )^{-7/2}
  \sigma v,
  \\
  &\simeq
  \frac{ 2 \pi^2 \rho_I(t) \Gamma_I }{p_0^{7/2} T^{3/2}(t)}
  \qty( \nu_{s_1} R_{s_2}^{\rm (asym)} +
\nu_{s_2} R_{s_1}^{\rm (asym)} )
\tilde{f}_{\rm tot}^{\rm (asym)}
  \frac{1}{4\pi^4}
  \alpha_{\chi}^2
  \frac{15 \sqrt{\pi}}{36} \frac{(T p_0)^{7/2} }{m_\chi^3},
  \\
  &= \frac{15 \alpha_{\chi}^2}{72 \pi^{3/2}}
  \frac{ \rho_I(t) \Gamma_I T^2(t)}{m_\chi^3} \qty( \nu_{s_1} R_{s_2}^{\rm (asym)} +
\nu_{s_2} R_{s_1}^{\rm (asym)} ) \tilde{f}_{\rm tot}^{\rm (asym)},
  \label{eq:boltzmannchi}
\end{align}
where we assume $p_0 T \gg m_\chi^2$ for simplicity.

To solve \eqref{eq:boltzmannchi}, we need the time-dependence of $T(t)$, which obeys
\begin{align}
 &\frac{\dd}{\dd t} \rho_r + 4 H \rho_r = \Gamma_I \rho_I,
 \\
 &H(t) = \sqrt{\frac{\rho_I + \rho_r}{3 \Mpl^2}},
 \\
 &\rho_r = \frac{g_* \pi^2}{30} T^4.
\end{align}
The time-dependence of $\rho_I$ is given by \eqref{eq:nI} with $\rho_I = m_I n_I$.
Integrating \eqref{eq:boltzmannchi} over $t$ in these equations,
we obtain
\begin{align}
 \frac{\rho_\chi}{s}
 &\simeq 4.4 \times
 \frac{45}{2 \pi^2 g_{*s}}
 \frac{15 \alpha_{\chi}^2}{72 \pi^{3/2}}
  \frac{ \Gamma_I^{3/2}\Mpl^{3/2}}{m_\chi^2}
  \qty( \nu_{s_1} R_{s_2}^{\rm (asym)} +
\nu_{s_2} R_{s_1}^{\rm (asym)} )
\tilde{f}_{\rm tot}^{\rm (asym)},
  \\
  &\simeq 2.2 \times 10^{-2} \,
  \alpha_\chi^2 \frac{T_{\rm RH}^3}{m_\chi^2}
  \qty( \nu_{s_1} R_{s_2}^{\rm (asym)} +
\nu_{s_2} R_{s_1}^{\rm (asym)} )
\tilde{f}_{\rm tot}^{\rm (asym)},
\end{align}
where $\rho_\chi = m_\chi n_\chi$, $s$ is the entropy density, and $g_{*s}$ is the effective number of degrees of freedom of the entropy density.
In the final line, we assume $g_* = g_{*s} = 106.75$.
This is consistent with the order of the estimated result in Ref.~\cite{Harigaya:2014waa}
up to a numerical factor, as expected.
In particular, the result is independent of $p_0$.
Note that the SM gauge coupling dependence is implicitly included in $\tilde{f}_{\rm tot}^{\rm (asym)}$ ($\sim \alpha_s^{-2}$) [see Eq.~\eqref{eq:tildeftotest}].

Let us substitute some numbers.
The factor of $\tilde{f}_{\rm tot}^{\rm (asym)}$ has a similar value for all $\si$ within an error of about $10\%$, so one may take its averaged value $\tilde{f}_{\rm tot, \si}^{\rm (asym)} \simeq 8.0$.
For example,
if $\chi$ are produced from the scattering of weak bosons,
we substitute $\nu_W = 6$, $R_W^{\rm (asym)} \simeq 0.0520$ and obtain
\begin{align}
 \frac{\rho_\chi}{s}
 &\simeq 0.11 \, \alpha_\chi^2 \frac{T_{\rm RH}^3}{m_\chi^2}.
 \label{eq:DMabundance}
\end{align}

Finally, we note that the result does not depend on the detail of the inflaton decay.
First of all, it is independent of the inflaton mass and the decay rate, as discussed in Ref.~\cite{Harigaya:2014waa}.
Moreover,
as we discuss in Sec.~\ref{sec:analytic} and show in Tabs.~\ref{tab:1}, \ref{tab:3}, and \ref{tab:4},
the asymptotic value of distributions in the scaling regime $R_s^{\rm (asym)} \tilde{f}_{\rm tot}^{\rm (asym)}$ is independent of the injected particle $\si$.
Our result is therefore almost independent of the high-energy physics even though a high-energy particle is injected from the heavy-particle decay.

%%%%%%%%%%%%%
\section{Discussion and conclusions}
\label{sec:conclusion}
%%%%%%%%%%%%%

We have investigated
the thermalization of high-energy SM particles in an ambient thermal plasma. This is realized, \textit{e.g.}, during the reheating after inflation, where an inflaton decays into SM particles with energy of the order of the inflaton mass.
All the relevant terms for splitting in the Boltzmann equations for all SM particles were taken into account, including the Abelian and non-Abelian gauge interactions and the top Yukawa interaction with the Higgs field.
The Boltzmann equations were numerically solved with a constant delta-function source term at $p = p_0$ with $p_0 \gg T$.
The distribution of particles scales as $p^{-7/2}$ for $p \ll p_0$, as expected from the scaling property of the splitting rate.

We have analytically calculated the asymptotic behavior of distributions at $p \approx p_0$ and $p \ll p_0$.
The first one represents the way in which the primary particles lose their energy, and the secondary and subsequent particles are produced during the initial thermalization.
The asymptotic behavior of distribution for primary particles at $p \approx p_0$ is proportional to $(p_0 - p)^{-1/2}$ for all SM particles other than the right-handed electron $e_f$ and hypercharge gauge boson $B$.
If the primary particle is $e_f$ or $B$, specific attention is required, where the distribution is proportional to a delta function. This is because the interaction of these particles is only the Abelian gauge, and the self-splitting rate is strongly suppressed.
Therefore, the particle distribution is not smeared from a delta-function source term.
Furthermore, we have shown that the asymptotic behavior of distributions at $p \ll p_0$ is independent of the injected primary particles.
In particular, the number density is dominated by gluons after a sufficiently large number of splittings occurs from any primary particle.
This is a strong prediction for the thermalization of high-energy SM particles.

Our results are useful when considering non-thermal production of DM during thermalization.
Since the distribution of SM particles is proportional to $p^{-7/2}$ at small $p$, a large amount of DM can be produced from scattering between the cascading particles and the thermal plasma.
The abundance of DM can be calculated using the result of distributions for a given energy $p$.
In particular, if the energy scale of the DM production process is much lower than the initial energy of the primary particles,
the scaling solution for the distributions can be used, and the result is independent of the way the primary particles are produced.
Furthermore, this result provides a definite prediction of DM abundance even though the process is highly non-thermal.
Moreover, there are several proposals for the non-thermal production of lepton asymmetry during the reheating era~\cite{Hamada:2015xva, Hamada:2018epb, Asaka:2019ocw}, for which the detailed thermalization process investigated in this paper may be important for calculating the produced lepton asymmetry.

We note that our results have broad applications.
Our assumptions are as follows:
i) Some SM particles are produced with a large energy $p_0$ and a small distribution $f_{\si} (p_0) \ll 1$,
ii) An ambient thermal plasma is present with a temperature $T$ ($\ll p_0$),
and
iii) The splitting rate is significantly greater than the Hubble expansion rate.
In several situations, these conditions are satisfied.
For example, if reheating is considered via the perturbative decay of inflatons, these conditions are satisfied after the temperature of the Universe reaches its maximal value~\cite{Mukaida:2015ria}.
The conditions are also satisfied at the last stage of reheating, in which case the temperature is equal to the reheating temperature.
For the decay of a long-lived heavy particle or extended object, our analysis is applicable to both cases where it dominates and does not dominate the Universe.

{\it Note added in v.3:}
We have corrected
typos in our numerical code
regarding $\mu_\perp^2$
by a factor of $\sqrt{4\pi}$ and the renormalization group running of gauge coupling constants especially in $m_{D,a}^2$.
This changes our results of Eqs.~\eqref{eq:gammatilde1},
\eqref{eq:gammatilde2},
\eqref{eq:gammatilde3},
\eqref{eq:gammatildeB1},
\eqref{eq:gammatildeB2},
\eqref{eq:gammatildeB3},
(\refeq{eq:constrainte}-\refeq{eq:constraintgamma}),
\eqref{eq:DMabundance},
and $\tilde{f}_{\rm tot,s_{\rm inj}}^{\rm (asym)}$ in Table~\ref{tab:3} and \ref{tab:4}
by a factor of about $\sqrt{4\pi}$.
Figures~\ref{fig:f-1}, \ref{fig:f-2} and \ref{fig:p0} are similarly rescaled.
This also changes $p^{(s_{\rm inj})}_{\rm asym}/p_0$, $p^{(s_{\rm inj})}_{\rm asym'}/p_0$ in Table~\ref{tab:3} and \ref{tab:4}.
Table~\ref{tab:1}
is changed within a few percent.
Figures~\ref{fig:R-1} and \ref{fig:R-2} are also slightly changed.

\section*{Acknowledgments}
K.\,M.\, was supported by MEXT Leading Initiative for Excellent Young Researchers Grant No.\ JPMXS0320200430,
and by JSPS KAKENHI Grant No.\ 	JP22K14044.
M.\,Y.\ was supported by the Leading Initiative for Excellent Young Researchers,
%Ministry of Education, Culture, Sports, Science and Technology (MEXT),
MEXT, Japan, and by JSPS KAKENHI Grant No.\ JP20H05851 and JP21K13910.

\appendix

%%%%%%%%%%%%%
\section{Boltzmann equations for the SM}
\label{sec:appendixA}
%%%%%%%%%%%%%

In this Appendix, we write the Boltzmann equations for the SM particles. Except for the top Yukawa, the Yukawa interactions can be neglected.
For notational simplicity, the argument for $\gamma_i$ and $f_s$ is omitted, which can be easily read from the general form of \eqref{eq:boltzmann} or the following:
\begin{align}
  \frac{\partial }{\partial t} f_s (p,t)
  &= - \frac{(2\pi)^3}{p^2 \nu_s}  \sum_{s',s''}
  \int_0^p  \dd k \,
    \gamma_{s \leftrightarrow s's''} \bigl(p; k, p-k \bigr) \,
    f_s(p)
    +
    \frac{(2\pi)^3}{p^2 \nu_s} \sum_{s',s''}
  \int_0^\infty \dd k \,
    \gamma_{s' \leftrightarrow s s''} \bigl(p+k; p, k \bigr) \,
    f_{s'}(p+k)
    \nn
    &\qquad + ({\rm source \ term}).
\end{align}

For the gluons,
\begin{align}
  \frac{\partial }{\partial t} f_g (p,t)
  &= - \frac{(2\pi)^3}{p^2 \nu_g}
  \int_0^p  \dd k \,
    \qty[ \gamma_{g \leftrightarrow gg} + \sum_f \qty(  \gamma_{g \leftrightarrow u_f \bar{u}_f} + \gamma_{g \leftrightarrow d_f \bar{d}_f} + 2 \gamma_{g \leftrightarrow Q_f \bar{Q}_f} ) ]
    f_g
    \nn
    &+
    \frac{(2\pi)^3}{p^2 \nu_g}
  \int_0^\infty \dd k \,
    \qty[
    2 \gamma_{g \leftrightarrow gg} f_g
    + \sum_f \qty( \gamma_{u_f \leftrightarrow g u_f} f_{u_f}
    + \gamma_{d_f \leftrightarrow g d_f} f_{d_f}
    + 2 \gamma_{Q_f \leftrightarrow gQ_f} f_{Q_f}
    ) ]
    .
    \label{eq:boltzman_g}
\end{align}
Similarly,
\begin{align}
  \frac{\partial }{\partial t} f_W (p,t)
  &= - \frac{(2\pi)^3}{p^2 \nu_W}
  \int_0^p  \dd k \,
    \qty[ \gamma_{W \leftrightarrow WW} + \sum_f \qty(  3 \gamma_{W \leftrightarrow Q_f \bar{Q}_f}
    + \gamma_{W \leftrightarrow L_f \bar{L}_f} )
    + \gamma_{W \leftrightarrow \phi \phi^*}      ]
    f_W
    \nn
    &+
    \frac{(2\pi)^3}{p^2 \nu_W}
  \int_0^\infty \dd k \,
    \qty[
    2 \gamma_{W \leftrightarrow WW} f_W
    + \sum_f \qty( 3 \gamma_{Q_f \leftrightarrow W Q_f} f_{Q_f}
    + \gamma_{L_f \leftrightarrow W L_f} f_{L_f} )
    + \gamma_{\phi \leftrightarrow W \phi} f_{\phi}
    ]
    ,
    \\
      \frac{\partial }{\partial t} f_B (p,t)
  &= - \frac{(2\pi)^3}{p^2 \nu_B}
  \int_0^p  \dd k \,
    \qty[ \sum_f \qty(
    3 \gamma_{B \leftrightarrow u_f \bar{u}_f}
    + 3 \gamma_{B \leftrightarrow d_f \bar{d}_f}
    + 6 \gamma_{B \leftrightarrow Q_f \bar{Q}_f}
    + \gamma_{B \leftrightarrow e_f \bar{e}_f}
    + 2 \gamma_{B \leftrightarrow L_f \bar{L}_f}
    )
    + 2 \gamma_{B \leftrightarrow \phi \phi^*}
    ]
    f_B
    \nn
    &+
    \frac{(2\pi)^3}{p^2 \nu_B}
  \int_0^\infty \dd k \,
    \qty[
     \sum_f
     \qty(
     3 \gamma_{u_f \leftrightarrow B u_f} f_{u_f}
     + 3 \gamma_{d_f \leftrightarrow B d_f} f_{d_f}
     + 6 \gamma_{Q_f \leftrightarrow B Q_f} f_{Q_f}
    + \gamma_{e_f \leftrightarrow B e_f} f_{e_f}
    + 2 \gamma_{L_f \leftrightarrow B L_f} f_{L_f}
    )
    + 2 \gamma_{\phi \leftrightarrow B \phi} f_{\phi}
    ]
    .
    \label{eq:boltzmann-photon}
\end{align}
For quarks, we have
\begin{align}
  \frac{\partial }{\partial t} f_{u_f} (p,t)
  &= - \frac{(2\pi)^3}{p^2 \nu_{u_f}}
  \int_0^p  \dd k \,
    \qty[  \gamma_{u_f \leftrightarrow g u_f}
    + 3 \gamma_{u_f \leftrightarrow B u_f}  ]
    f_{u_f}
    \nn
    &+
    \frac{(2\pi)^3}{p^2 \nu_{u_f}}
  \int_0^\infty \dd k \,
    \qty[
     2 \gamma_{g \leftrightarrow u_f \bar{u}_f} f_g
    +  6 \gamma_{B \leftrightarrow u_f \bar{u}_f} f_B
    + \qty( \gamma_{u_f \leftrightarrow u_f g}
    + 3 \gamma_{u_f \leftrightarrow u_f B} ) f_{u_f}
    ] \,,
    \label{eq:boltzmannuf}
\end{align}
for $f= 1,2$
and similar for $d_f$ with $f=1,2,3$. Here we included a factor of $2$ for pair creation since we involved anti-particle for $\nu_s$.
For the left-handed quarks, we have
\begin{align}
  \frac{\partial }{\partial t} f_{Q_f} (p,t)
  &= - \frac{(2\pi)^3}{p^2 \nu_{Q_f}}
  \int_0^p  \dd k \,
    \qty[ \qty( 2 \gamma_{Q_f \leftrightarrow g Q_f}
    + 3 \gamma_{Q_f \leftrightarrow W Q_f}
    + 6 \gamma_{Q_f \leftrightarrow B Q_f}  ) ]
    f_{Q_f}
    \nn
    &+
    \frac{(2\pi)^3}{p^2 \nu_{Q_f}}
  \int_0^\infty \dd k \,
    \qty[
     4 \gamma_{g \leftrightarrow Q_f \bar{Q}_f} f_g
    + 6 \gamma_{W \leftrightarrow Q_f \bar{u}_f} f_W
    +  12 \gamma_{B \leftrightarrow Q_f \bar{Q}_f} f_B
    + \qty( 2 \gamma_{Q_f \leftrightarrow Q_f g}
    + 3 \gamma_{Q_f \leftrightarrow Q_f W}
    + 6 \gamma_{Q_f \leftrightarrow Q_f B} ) f_{Q_f}
    ]
    \label{eq:boltzmannQf}
\end{align}
for $f= 1,2$.
For the top quark,
we should add contribution from the Yukawa interaction, such as
\begin{align}
  \frac{\partial }{\partial t} f_{u_3} (p,t)
  &= - \frac{(2\pi)^3}{p^2 \nu_{u_3}}
  \int_0^p  \dd k \,
    \qty[ \dots + 3 \gamma_{u_3 \leftrightarrow \phi Q_3}  ]
    f_{u_3}
    +
    \frac{(2\pi)^3}{p^2 \nu_{u_3}}
  \int_0^\infty \dd k \,
    \qty[
    \dots
    + 6 \gamma_{\phi \leftrightarrow u_3 \bar{Q}_3} f_{\phi}
    + 3 \gamma_{Q_3 \leftrightarrow u_3 \phi^* } f_{Q_3}
    ],
\label{eq:f_u3}
    \\
      \frac{\partial }{\partial t} f_{Q_3} (p,t)
  &= - \frac{(2\pi)^3}{p^2 \nu_{Q_3}}
  \int_0^p  \dd k \,
    \qty[ \dots + 3 \gamma_{Q_3 \leftrightarrow \phi^* u_3}  ]
    f_{Q_3}
    +
    \frac{(2\pi)^3}{p^2 \nu_{Q_3}}
  \int_0^\infty \dd k \,
    \qty[
    \dots
    + 6 \gamma_{\phi \leftrightarrow \bar{Q}_3 u_3 } f_{\phi}
    + 3 \gamma_{u_3 \leftrightarrow Q_3 \phi} f_{u_3}
    ]
    .
\label{eq:f_Q3}
\end{align}
For the leptons, we have
\begin{align}
  \frac{\partial }{\partial t} f_{e_f} (p,t)
  &= - \frac{(2\pi)^3}{p^2 \nu_{e_f}}
  \int_0^p  \dd k \,
    \qty[
     \gamma_{e_f \leftrightarrow B e_f}  ]
    f_{e_f}
    +
    \frac{(2\pi)^3}{p^2 \nu_{e_f}}
  \int_0^\infty \dd k \,
    \qty[
     2 \gamma_{B \leftrightarrow e_f \bar{e}_f} f_B
    + \gamma_{e_f \leftrightarrow e_f B} f_{e_f}
    ]
    ,
    \label{eq:boltzmannef}
    \\
  \frac{\partial }{\partial t} f_{L_f} (p,t)
  &= - \frac{(2\pi)^3}{p^2 \nu_{L_f}}
  \int_0^p  \dd k \,
    \qty[
    \gamma_{L_f \leftrightarrow W L_f}
    + 2 \gamma_{L_f \leftrightarrow B L_f} ]
    f_{L_f}
    \nn
    &+
    \frac{(2\pi)^3}{p^2 \nu_{L_f}}
  \int_0^\infty \dd k \,
    \qty[
      2 \gamma_{W \leftrightarrow L_f \bar{L}_f} f_W
    +  4 \gamma_{B \leftrightarrow L_f \bar{L}_f} f_B
    + \qty( \gamma_{L_f \leftrightarrow L_f W}
    + 2 \gamma_{L_f \leftrightarrow L_f B} ) f_{L_f}
    ] \,,
    \label{eq:boltzmannLf}
\end{align}
for $f= 1,2,3$.
The Higgs field obeys
\begin{align}
  \frac{\partial }{\partial t} f_{\phi} (p,t)
  &= - \frac{(2\pi)^3}{p^2 \nu_{\phi}}
  \int_0^p  \dd k \,
    \qty[
      \gamma_{\phi \leftrightarrow W \phi}
     + 2 \gamma_{\phi \leftrightarrow B \phi}
     + 6 \gamma_{\phi \leftrightarrow u_3 \bar{Q}_3}  ]
    f_{\phi}
    \nn
    &+
    \frac{(2\pi)^3}{p^2 \nu_{\phi}}
  \int_0^\infty \dd k \,
    \qty[
     2 \gamma_{W \leftrightarrow \phi \phi^*} f_W
     + 4 \gamma_{B \leftrightarrow \phi \phi^*} f_B
     + \qty( \gamma_{\phi \leftrightarrow \phi W}
     + 2 \gamma_{\phi \leftrightarrow \phi B} ) f_\phi
    + 3 \gamma_{u_3 \leftrightarrow \phi Q_3} f_{u_3}
    + 3 \gamma_{Q_3 \leftrightarrow \phi^* u_3} f_{Q_3}
    ]
    .
\label{eq:f_phi}
\end{align}

\section{Boundary condition and asymptotic behavior at $p \approx p_0$}
\label{sec:appendixB}

In this Appendix, the asymptotic behaviors of distributions at $p \approx p_0$ are calculated, which are useful to impose boundary conditions for distributions in numerical calculations.
The asymptotic behavior is different for the primary and other particles.

\subsection{Primary particles}

\subsubsection{Case with primary gluon}

First, we consider the case in which gluons were injected at $p = p_0$ with a delta-function source term.
The stationary equation for the gluon distribution is similar to the following:
\begin{align}
  - 2 \int_0^{p/2}  \dd k \,
    \gamma_{g \leftrightarrow g g} \bigl(p; k, p-k \bigr) \,
    f_g(p)
    +
  \int_0^\infty \dd k \,
    2 \gamma_{g \leftrightarrow g g} \bigl(p+k; p, k \bigr) \,
    f_g(p+k)
    + \frac{p^2 }{(2\pi)^3} p_0^{1/2} T^{3/2} \tilde{\Gamma} \delta (p - p_0) = 0,
\end{align}
where we include the source term ($= p_0^{1/2} T^{3/2} \Br \tilde{\Gamma} \delta (p - p_0)$)
and use $\Br = 1/\nu_g$.
Here, we neglected terms associated with quarks because the secondary particles are subdominant at $p \approx p_0$, and no IR divergence is observed in the pair production of quarks.
Defining $x \equiv k / p$ and $x_{\rm max} \equiv p_0 / p$, we can rewrite it as
\begin{align}
  - \tilde{\gamma}_{g \leftrightarrow g g} \qty[ \int_0^{1/2}  \frac{\dd x}{x^{3/2}} \,
    f_g(p)
    -
  \int_0^{x_{\rm max}-1} \frac{\dd x}{x^{3/2}} \,
    f_g(p(1+x))
    ] + \frac{p}{(2\pi)^3} \tilde{\Gamma} \delta (x_{\rm max}-1) \approx 0,
\label {eq:gluonasym}
\end{align}
where we approximate
$\gamma_{g \leftrightarrow g g} \bigl(p; k, p-k \bigr) \simeq
\gamma_{g \leftrightarrow g g} \bigl(p+k; p, k \bigr) \simeq p^{1/2} T^{3/2} \tilde{\gamma}_{g \leftrightarrow g g} x^{-3/2}/2$ with a dimensionless constant $\tilde{\gamma}_{g \leftrightarrow g g}$ for $x \ll 1$.
Although this approximation is not satisfactory for all integral domain in the first term, the contribution near $x \sim 1$ is subdominant and is not significant for our discussion.
Now we make an ansatz $f_g(p) = C_g ((p_0 - p)/p_0)^{-1/2 + \epsilon}$
with $C_g$ and $\epsilon$ being constants.
Then we can perform the integral and obtain
\begin{align}
  - \frac{2 \pi \epsilon}{\sqrt{x_{\rm max} - 1}} \tilde{\gamma}_{g \leftrightarrow g g} \, C_g ((p_0 - p)/p_0)^{-1/2 + \epsilon}
  + \frac{p}{(2\pi)^3} \tilde{\Gamma} \delta (x_{\rm max}-1) \approx 0,
\label {eq:bc1}
\end{align}
at the leading order of $x_{\rm max} -1 \ll 1$ for $\epsilon \to 0$.
Here we used
\begin{align}
- \int_0^{1/2}  \frac{\dd x}{x^{3/2}} \,
+
  \int_0^{x_{\rm max}-1} \frac{\dd x}{x^{3/2}} \, \qty( \frac{x_{\rm max} -1 - x}{x_{\rm max} - 1} )^{-1/2 + \epsilon}
  &\simeq - \frac{1}{\sqrt{x_{\rm max} - 1}}
  \frac{2 \sqrt{\pi}\, \Gamma(1/2 + \epsilon)}{\Gamma(\epsilon)}
  \\
  &\simeq - \frac{2 \pi \epsilon}{\sqrt{x_{\rm max} - 1}}
  \quad \text{for} \ \epsilon \to 0,
\end{align}
where we take a limit of $x_{\rm max} -1 \ll 1$ in the first line.
We can check that \eqref{eq:bc1} is consistent by taking an integral over $p$ from $p_0( 1- \delta)$ to $p_0$ with a small $\delta$ and taking
a limit of $\epsilon \to 0$ because the first term provides
\begin{align}
  &- 2 \pi \epsilon \tilde{\gamma}_{g \leftrightarrow g g} \, C_g
  \int_{p_0(1-\delta)}^{p_0} \dd p \sqrt{\frac{p}{p_0 - p}} \qty( \frac{p_0 - p}{p_0} )^{-1/2 + \epsilon}
  \label{eq:deltafunct}
  \\
  &= - 2 \pi \tilde{\gamma}_{g \leftrightarrow g g} \, C_g p_0
  \qty( \delta^\epsilon - 0 )
  \\
  &\to - 2 \pi \tilde{\gamma}_{g \leftrightarrow g g} \, C_g p_0
  \quad \text{for} \ \epsilon \to 0.
\end{align}
Thus, we can determine $C_g$ to satisfy \eqref{eq:bc1} such that%
\footnote{
One may think that
the delta function provides $1/2$ for the integral over $p$ from $p_0( 1- \delta)$ to $p_0$.
However,
as we discussed in footnote~\ref{footnote1},
it should give a factor of unity for the integral over the line segment of $p \le p_0$ in our definition of delta function or source term.
}
\begin{equation}
 C_g = \frac{\tilde{\Gamma}}{(2\pi)^4 \tilde{\gamma}_{g \leftrightarrow g g}}.
\end{equation}
In summary, we obtain the asymptotic behavior for $f_g$ such that
\begin{equation}
 f_g(p) \approx \frac{\tilde{\Gamma}}{ (2\pi)^4 \tilde{\gamma}_{g \leftrightarrow g g}}  \qty( \frac{p_0 - p}{p_0} )^{-1/2}
\end{equation}
for $p \approx p_0$.

\subsubsection{Case with primary $L_f, u_f, d_f, Q_f, \phi, W$}

A similar result with gluon is observed for primary particles with IR divergent splitting rates, such as $\si = L_f$, $u_{f'}$, $u_3$, $d_f$, $Q_{f'}$, $Q_3$, $\phi$, and $W$, once we replace $\tilde{\gamma}_{g \leftrightarrow g g}$ appropriately.
Here we summarize the results:
\begin{equation}
 f_{s_1} (p) \approx
  \frac{\tilde{\Gamma}}{(2\pi)^4  \sum_{a'} \tilde{\gamma}_{s_{1} \to g_{a'} s_{1}}
 }  \qty( \frac{p_0 - p}{p_0} )^{-1/2}
\end{equation}
for $s_1 = L_f, u_{f'}, u_3, d_f, Q_{f'}, Q_3, \phi, g, W$.

Here we explicitly write $\tilde{\gamma}_{s_{1} \to g_a s_{1}}$ as:
\begin{align}
 &\sum_{a'} \tilde{\gamma}_{g_a \to g_{a'} g_a} = \frac{x^{3/2}}{p^{1/2} T^{3/2}} \qty( \frac{\nu_{g_a}}{ d_{A_a}^{(a)}} \gamma_{g_a \leftrightarrow g_a g_a} )
 \quad {\rm for} \ g_a = (g, W)
 \label{eq:tildegtogg}
 \\
 &\sum_a \tilde{\gamma}_{s_f \to g_a s_f} = \frac{x^{3/2}}{p^{1/2} T^{3/2}} \qty( \sum_a
 \frac{\nu_{s_f}}{2 d_{F_s}^{(a)}} \gamma_{s_f \leftrightarrow a s_f} )
 \quad {\rm for} \ s_f = (L_f, u_{f'}, u_3, d_f, Q_{f'}, Q_3),
 \label{eq:tildeqtogq}
 \\
 &\sum_a \tilde{\gamma}_{\phi \to g_a \phi} = \frac{x^{3/2}}{p^{1/2} T^{3/2}} \qty( \sum_a
 \frac{\nu_\phi}{2 d_\phi^{(a)}} \gamma_{\phi \leftrightarrow g_a \phi} )
 \label{eq:tildephitogphi}
\end{align}
for $x \ll 1$, where we should evaluate at $p=p_0$.
Here, $a$ can run from 1 to 3 in the summation. However, the contribution from $a = 1$ vanishes in the limit of $x \to 0$, and this is because the soft $B$ boson emission is suppressed by an additional factor of $x$.
This behaviour implies that specific attention is paid to the right-handed lepton, which has the gauge interaction only with $B$ as explained later.

As a reference, we obtain
\begin{align}
 (2\pi)^4 \sum_{a'} \tilde{\gamma}_{s_{1} \to g_{a'} s_{1}}
 &\simeq
 (
 0.053, \
 0.23, \
 0.23, \
 0.23, \
 0.62, \
 0.62, \
 0.053, \
 4.9, \
 0.42
 )
 \nonumber\\
 &{\rm for} \ s_1 = (L_f, \ u_{f'}, \ u_3, \ d_f, \ Q_{f'}, \ Q_3, \ \phi, \ g, \ W),
 \label{eq:gammatildeB1}
\end{align}
respectively.
Here we assumed $p = p_0 = 10^{12} T = 10^{15} \GeV$, though the dependence on $p$ is only logarithmic through the running of coupling constants.

\subsubsection{Case with primary $B$}

The cases with $\si = e_f$ and $B$ have qualitatively different distributions at $p \approx p_0$. This is because they experience only Abelian gauge interactions that lead to a different scaling for the splitting rate [see Eq.~\eqref{eq:split_func_rough}].

If the primary particle is an Abelian gauge boson $B$, it splits into other particles and does not undergo soft-dominated splitting processes.
The relevant part of its Boltzmann equation can be written as
\begin{align}
  - \sum_{s_f} \tilde{\gamma}_{B \to s_f \bar{s}_f}
 \int_0^{1}  \dd x \frac{x^2 + (1-x)^2}{x^{1/2}(1-x)^{1/2}} \,
    f_B(p)
  - \tilde{\gamma}_{B \to \phi \phi^*}
 \int_0^{1}  \dd x \frac{2 \sqrt{x(1-x)}}{1} \,
    f_B(p)
+ \frac{p_0}{(2\pi)^3} \tilde{\Gamma}  \delta(p-p_0)
    \approx 0,
    \label{eq:boltzmanphoton}
\end{align}
where
\begin{align}
 &\sum_{s_f} \tilde{\gamma}_{B \to s_f \bar{s}_f}
 \simeq \frac{1}{p^{1/2} T^{3/2}}\qty[ \frac{x^2 + (1-x)^2}{x^{1/2}(1-x)^{1/2}} ]^{-1}
 \qty( \sum_{s_f} \frac{\nu_{s_f}}{2 d_{F_s}^{(1)}} \gamma_{B  \leftrightarrow s_f \bar{s}_f} )
 \label{tildeBtoqq}
\\
 &\tilde{\gamma}_{B \to \phi \phi^*}
 = \frac{1}{p^{1/2} T^{3/2}}
 \qty( \frac{2\sqrt{x(1-x)}}{1} )^{-1}  \qty( \frac{\nu_{\phi}}{2 d_{\phi}^{(1)}} \gamma_{B  \leftrightarrow \phi \phi^*} )
 \label{tildeBtophiphi}
\end{align}
for $x \ll 1$.
Here, the summation for $s_f$ ($= e_f, L_f, u_{f'}, u_3, d_f, Q_{f'}, Q_3$) included all flavors.
The integrals do not have the IR divergence and
can be performed analytically. We then obtain
the delta function distribution for $f_B(p)$ as:
\begin{equation}
 f_B(p) = C_B' \delta (p - p_0) + C_B \qty( \frac{p_0 - p}{p_0} )^{m'},
\end{equation}
where we include a subdominant component that is determined later.
Here, $C_B'$ is determined to satisfy \eqref{eq:boltzmanphoton} such that
\begin{equation}
 C_B'
    = \frac{p_0}{(\pi/4) (2\pi)^3} \frac{\tilde{\Gamma}}{3 \sum_{s_f} \tilde{\gamma}_{B \to s_f s_f} + \tilde{\gamma}_{B \to \phi \phi} }.
\end{equation}
As a reference, we obtain
\begin{align}
 (\pi/4) (2\pi)^3 \qty[ 3 \sum_{s_f} \tilde{\gamma}_{B \to s_f s_f} + \tilde{\gamma}_{B \to \phi \phi} ]
 &\simeq
 0.041,
 \label{eq:gammatildeB2}
\end{align}
where we assume $p = p_0 = 10^{12} T = 10^{15} \GeV$ though the dependence on $p$ is only logarithmic.

From the delta-function distribution of $B$,
other particles are produced from the pair production processes.
This is discussed in the next section.
The subdominant part of the distribution for photon is determined by emission from $e_f$ [see Eq.~\eqref{eq:formula3}].

\subsubsection{Case with primary $e_f$}

Similar to the $B$ emission, $e_f$ experiences only Abelian gauge interaction.
Contrary to the non-Abelian case,
the soft-photon emission rate is more suppressed by the LPM effect by a factor of $k/p$, which removes the IR divergence.
As a result, the delta-function distribution from the source term cannot be softened and
$e_f$ has the delta function distribution plus a subdominant component such that
\begin{equation}
 f_{e_f} (p) = C_{e_f}' \delta (p - p_0) + C_{e_f} \qty( \frac{p_0 - p}{p_0} )^{m'}.
\end{equation}
Here,
$C_{e_f}'$ satisfies
\begin{equation}
 C_{e_f}'
 \int_0^p  \dd k \,
    \gamma_{e_f \leftrightarrow B e_f}  \bigl(p; k, p-k \bigr)
    = \frac{p_0^2}{ (2\pi)^3} p_0^{1/2} T^{3/2}\tilde{\Gamma}.
\end{equation}
This equation can be approximated by
\begin{align}
  C_{e_f}' \tilde{\gamma}_{e_f \to B e_f}
  \int_0^{1}  \dd x \frac{1 + (1-x)^2}{x^{1/2}(1-x)^{1/2}} \,
  \approx \frac{p_0}{ (2\pi)^3} \tilde{\Gamma},
\end{align}
where
\begin{align}
 \tilde{\gamma}_{e_f \to B e_f} =
 \frac{1}{p^{1/2} T^{3/2}} \qty[ \frac{1 + (1-x)^2}{x^{1/2}(1-x)^{1/2}} ]^{-1}
 \gamma_{e_f  \leftrightarrow B e_f}
 \label{eq:tildeetoBe}
\end{align}
for $x \ll 1$.
This provides
\begin{align}
  C_{e_f}'
  \approx \frac{p_0}{ (2\pi)^3}
  \frac{8}{11 \pi} \frac{ \tilde{\Gamma}}{\tilde{\gamma}_{e_f \to B e_f}}.
\end{align}
As a reference, we obtain
\begin{align}
 (2\pi)^3 \frac{11 \pi}{8} \tilde{\gamma}_{e_f \to B e_f}
 &\simeq
 1.3 \times 10^{-3},
 \label{eq:gammatildeB3}
\end{align}
where we assume $p = p_0 = 10^{12} T = 10^{15} \GeV$ though the dependence on $p$ is only logarithmic.

The subdominant part is determined from
\begin{align}
&-  \int_0^p  \dd k \,
     \gamma_{e_f \leftrightarrow B e_f}  \bigl(p; k, p-k \bigr) \,
    f_{e_f} (p)
    +
  \int_0^\infty \dd k \,
    \gamma_{e_f \leftrightarrow e_f B} \bigl(p+k; p, k \bigr) \, f_{e_f} (p+k)
    = 0.
\end{align}
Substituting the distribution into this equation, we obtain
\begin{align}
 &C_{e_f} =
\frac{8}{11\pi p_0}
C_{e_f}'
 \\
 &m' = -1/2.
\end{align}

\subsection{Secondary particles}

Next, we calculate the asymptotic behavior of $p \approx p_0$ for secondary (and subsequent) particles that are produced from the primary particle.

First, we consider a toy model to represent the basic phenomenon.
If a fermion $s_{n+1}$ is produced from a gauge boson $s_{n}$ with a known distribution $f_{s_n}$ with the following stationary Boltzmann equation:
\begin{align}
&-
  \int_0^p  \dd k \,
      \gamma_{s_{n+1} \leftrightarrow s' s_{n+1}}     \bigl(p; k, p-k \bigr) \,
    f_{s_{n+1}} (p)
    \nn
    &\qquad +
  \int_0^{p_0-p} \dd k \,
    \qty[
     2 \gamma_{s_{n} \leftrightarrow s_{n+1} s_{n+1}} \bigl(p+k; p, k \bigr) \,
     f_{s_{n}} (p+k)
    + \gamma_{s_{n+1} \leftrightarrow s_{n+1} s' }
    \bigl(p+k; p, k \bigr) \,
 f_{s_{n+1}} (p+k)
    ]
    = 0
    ,
    \label{eq:toy}
\end{align}
where $s'$ is a non-Abelian gauge boson or a Higgs boson, which may or may not be $s_n$.
We define $x \equiv k / p$ and $x_{\rm max} \equiv p_0 / p$ and rewrite the equation as
\begin{align}
  - \tilde{\gamma}_{s_{n+1} \leftrightarrow s' s_{n+1}} \qty[ \int_0^{1}  \frac{\dd x}{x^{3/2}} \,
    f_{s_{n+1}}(p)
    -
  \int_0^{x_{\rm max}-1} \frac{\dd x}{x^{3/2}} \,
    f_{s_{n+1}}(p(1+x))
    ]
    +
    2 \tilde{\gamma}_{s_{n} \leftrightarrow s_{n+1} s_{n+1}} \int_0^{x_{\rm max}-1}  \frac{\dd x}{x^{1/2}} \,
    f_{s_{n}}(p(1+x))
    \approx 0,
\end{align}
where
$\gamma_{s_{n+1} \leftrightarrow s' s_{n+1}}     \bigl(p; k, p-k \bigr) \simeq
\gamma_{s_{n+1} \leftrightarrow s_{n+1} s' }     \bigl(p+k; p, k \bigr) \simeq p^{1/2} T^{3/2} \tilde{\gamma}_{s_{n+1} \leftrightarrow s' s_{n+1}}x^{-3/2}$
and
$\gamma_{s_{n} \leftrightarrow s_{n+1} s_{n+1}} \bigl(p+k; p, k \bigr) \simeq p^{1/2} T^{3/2} \tilde{\gamma}_{s_{n} \leftrightarrow s_{n+1} s_{n+1}} x^{-1/2}$
with a constant $\tilde{\gamma}_{s_{n+1} \leftrightarrow s' s_{n+1}}$
and $\tilde{\gamma}_{s_{n} \leftrightarrow s_{n+1} s_{n+1}}$
for $x \ll 1$.
Now,
suppose that $f_{s_n}$ has the form of $f_{s_n} = C_{s_n} \qty( (p_0 - p) / p_0 )^m$ with a constant $C_{s_n}$ and $m$.
Using an ansatz $f_{s_{n+1}} = C_{s_{n+1}} \qty( (p_0 - p) / p_0 )^{m'}$, we can perform the above integral and obtain
\begin{align}
  - 2 \sqrt{\pi} \frac{\Gamma(1 + m')}{ \Gamma(1/2+m')} C_{s_{n+1}} \tilde{\gamma}_{s_{n+1} \leftrightarrow s' s_{n+1}} \qty( (p_0 - p) / p_0 )^{m'-1/2}
    +
    2 \sqrt{\pi} \frac{\Gamma(1 + m)}{\Gamma(3/2+m)} C_{s_{n}} \tilde{\gamma}_{s_{n} \leftrightarrow s_{n+1} s_{n+1}} \qty( (p_0 - p) / p_0 )^{m+1/2}
    \approx 0
\end{align}
for a leading order of $x_{\rm max} - 1 \ll 1$.
Therefore, we should take
\begin{align}
 &C_{s_{n+1}} = \frac{1}{1+m}  \frac{\tilde{\gamma}_{s_{n} \leftrightarrow s_{n+1} s_{n+1}}}{\tilde{\gamma}_{s_{n+1} \leftrightarrow s' s_{n+1}}} C_{s_{n}}
 \\
 &m' = m + 1
\end{align}
for the distribution of particle species $s_{n+1}$.

The form of \eqref{eq:toy} corresponds to the splitting process from non-Abelian gauge bosons to fermions.
A similar discussion and conclusion holds for the emission of non-Abelian gauge bosons from fermions, by replacing
$\tilde{\gamma}_{s_{n} \leftrightarrow s_{n+1} s_{n+1}}$
and $\tilde{\gamma}_{s_{n+1} \leftrightarrow s' s_{n+1}}$ appropriately.
However, the one associated with Abelian gauge boson should be provided specific attention because of the absence of IR divergence as we will see in the next section.

\subsubsection{Case with $L_f, u_f, d_f, Q_f, \phi, g$, and $W$ production}

From the discussion above, the distribution of $s_{n+1}$ is determined by the following procedure: the second term of \eqref{eq:toy} is calculated from a known distribution $f_{s_n}$ and compared with the first and third terms calculated from the ansatz. We generically denote the stationary Boltzmann equation in the limit of $x_{\rm max} - 1 \ll 1$ such that
\begin{align}
  - \sum_a \tilde{\gamma}_{s_{n+1} \to g_a s_{n+1}}
  \qty[ \int_0^{1}  \frac{\dd x}{x^{3/2}} \,
    f_{s_{n+1}}(p)
    -
  \int_0^{x_{\rm max}-1} \frac{\dd x}{x^{3/2}} \,
    f_{s_{n+1}}(p(1+x))
    ]
    +
   \tilde{\gamma}_{s_{n} \to s_{n+1}} \int_0^{x_{\rm max}-1}  \frac{\dd x}{x^{b}} \,
    f_{s_{n}}(p(1+x))
    \approx 0
\end{align}
for $s_{n+1} = L_f, u_f, d_f, Q_f, \phi, g$, and $W$,
where $\tilde{\gamma}_{s_{n+1} \to g_a s_{n+1}}$ is given by
Eqs.~(\ref{eq:tildegtogg}), (\ref{eq:tildeqtogq}), and (\ref{eq:tildephitogphi}).
Here, $b = -1/2, 1/2$, or $3/2$ depending on the splitting,
and the explicit form of $\tilde{\gamma}_{s_{n} \to s_{n+1}}$ is given below.
Taking
$f_{s_{n+1}} = C_{s_{n+1}} \qty( (p_0 - p) / p_0 )^{m'}$, we obtain
\begin{align}
 &C_{s_{n+1}} = \frac{\Gamma(1/2 + m')}{2 \sqrt{\pi} \Gamma(1+m')} \frac{\Gamma(1-b) \Gamma (1+m)}{\Gamma(2-b+m)}  \frac{\tilde{\gamma}_{s_{n} \to s_{n+1}}}{\sum_a \tilde{\gamma}_{s_{n+1} \to g_a s_{n+1}}} C_{s_{n}}
 \\
 &m' = m + 3/2 - b
\end{align}
for $f_{s_n} = C_{s_n} \qty( (p_0 - p) / p_0 )^m$
or
\begin{align}
 &C_{s_{n+1}} = \frac{\Gamma(1/2 + m')}{2 \sqrt{\pi} \Gamma(1+m')}  \frac{\tilde{\gamma}_{s_{n} \to s_{n+1}}}{\sum_a \tilde{\gamma}_{s_{n+1} \to g_a s_{n+1}}} C_{s_{n}}'
 \\
 &m' = 1/2 - b
\end{align}
for $f_{s_n} = C_{s_n}' \delta (p - p_0)$.

Here we provide $\tilde{\gamma}_{s_{n} \to s_{n+1}}$ for each process:
\begin{align}
 &\tilde{\gamma}_{g_a \to s_f} = \frac{x^{1/2}}{p^{1/2} T^{3/2}}  \qty(
 2 \frac{\nu_{s_f}}{2 d_{F_s}^{(a)}} \gamma_{g_s \leftrightarrow s_f \bar{s}_f} )
 \quad {\rm with} \ b = 1/2
 \quad {\rm for} \ s_f = (e_f, L_f, u_{f'}, u_3, d_f, Q_{f'}, Q_3), \ \ g_a = (g, W, B)
 \label{eq:tildegatoqs}
 \\
 &\tilde{\gamma}_{s_f \to g_{a'}} = \frac{x^{1/2}}{p^{1/2} T^{3/2}} \qty(
 \frac{\nu_{s_f}}{2 d_{F_s}^{(a')}} \gamma_{s_f \leftrightarrow g_{a'} s_f} )
 \quad {\rm with} \ b = 1/2
 \quad {\rm for} \ s_f = (L_f, u_{f'}, u_3, d_f, Q_{f'}, Q_3), \ \ g_a = (g, W, B),
 \label{eq:tildeqstoga}
 \\
 &\tilde{\gamma}_{g_a \to \phi} = \frac{x^{-1/2}}{p^{1/2} T^{3/2}} \qty(
 2 \frac{\nu_{\phi}}{2 d_{\phi}^{(a)}} \gamma_{g_s \leftrightarrow \phi \phi^*} )
 \quad {\rm with} \ b = -1/2
 \quad {\rm for} \ g_a = (g, W, B)
 \\
 &\tilde{\gamma}_{\phi \to g_{a'}} = \frac{x^{-1/2}}{p^{1/2} T^{3/2}} \qty(
 \frac{\nu_{\phi}}{2 d_{\phi}^{(a')}} \gamma_{\phi \leftrightarrow g_{a'} \phi} )
 \quad {\rm with} \ b = -1/2
 \quad {\rm for} \ g_a = (g, W, B),
 \label{eq:tildephitoga}
  \\
 &\tilde{\gamma}_{s_f \to \phi} = \frac{x^{1/2}}{p^{1/2} T^{3/2}} \qty(
 3 \gamma_{s_f \leftrightarrow \phi s_{f'}} )
 \quad {\rm with} \ b = 1/2
 \quad {\rm for} \ (s_f, s_{f'}) = (u_3, Q_3) \ {\rm or} \  (Q_3, u_3)
 \\
 &\tilde{\gamma}_{\phi \to s_f} = \frac{x^{1/2}}{p^{1/2} T^{3/2}} \qty(
 6 \gamma_{\phi \leftrightarrow s_f s_{f'}} )
 \quad {\rm with} \ b = 1/2
 \quad {\rm for} \ (s_f, s_{f'}) = (u_3, Q_3) \ {\rm or} \  (Q_3, u_3)
  \\
 &\tilde{\gamma}_{s_f \to s_{f'}} = \frac{x^{-1/2}}{p^{1/2} T^{3/2}} \qty(
 3 \gamma_{s_f \leftrightarrow \phi s_{f'}} )
 \quad {\rm with} \ b = -1/2
 \quad {\rm for} \ (s_f, s_{f'}) = (u_3, Q_3) \ {\rm or} \  (Q_3, u_3)
\end{align}
for $x \ll 1$.
Generically, $s_f$ represents a single generation $f$. We may take a summation over $f$
in \eqref{eq:tildegatoqs} to obtain a total splitting rate
[see, \textit{e.g.}, Eq.~\eqref{eq:boltzman_g}].

The aforementioned results imply that the production of Higgs boson is suppressed by an additional power of $(p_0 - p)/p_0$ for $p \approx p_0$.
A similar suppression is obtained for the emission of $W$ from $\phi$
and the production of right/left-handed top quark from left/right-handed quark via the Yukawa interaction.

\subsubsection{Case with $B$ production}
The emission of photons has a qualitatively different behavior than that for non-Abelian gauge bosons.
For $s_{n+1} = B$, we generically write
\begin{align}
  - \sum_{s_f} \tilde{\gamma}_{B \to s_f \bar{s}_f}
 \int_0^{1}  \dd x \frac{x^2 + (1-x)^2}{x^{1/2}(1-x)^{1/2}} \,
    f_{B}(p)
  - \tilde{\gamma}_{B \to \phi \phi^*}
 \int_0^{1}  \dd x \frac{2 \sqrt{x(1-x)}}{1} \,
    f_{B}(p)
    +
    \tilde{\gamma}_{s_{n} \to B} \int_0^{x_{\rm max}-1}  \frac{\dd x}{x^{b}} \,
    f_{s_{n}}(p(1+x))
    \approx 0\,,
\end{align}
where $s_f$ represents the fermions.
Here,
$\tilde{\gamma}_{s_{n} \to B}$ is given by Eqs.~(\ref{eq:tildeqstoga}) and (\ref{eq:tildephitoga}), whereas $\sum_{s_f} \tilde{\gamma}_{B \to s_f s_f}$ and $\tilde{\gamma}_{B \to \phi \phi}$ are given by Eqs.~(\ref{tildeBtoqq}) and (\ref{tildeBtophiphi}).
Taking
$f_{B} = C_{B} \qty( (p_0 - p) / p_0 )^{m'}$, we obtain
\begin{align}
 &C_{B} =
\frac{4}{\pi} \frac{\Gamma(1-b) \Gamma (1+m)}{\Gamma(2-b+m)}  \frac{\tilde{\gamma}_{s_{n} \to B}}{3 \sum_{s_f} \tilde{\gamma}_{B \to s_f \bar{s}_f} + \tilde{\gamma}_{B \to \phi \phi^*} } C_{s_{n}}
 \\
 &m' = m + 1 - b
\end{align}
for $f_{s_n} = C_{s_n} \qty( (p_0 - p) / p_0 )^m$
or
\begin{align}
 &C_{B} =
 \frac{4}{\pi} \frac{\tilde{\gamma}_{s_{n} \to B}}{3 \sum_{s_f} \tilde{\gamma}_{B \to s_f \bar{s}_f} + \tilde{\gamma}_{B \to \phi \phi^*} }
C_{s_{n}}'
 \label{eq:formula3}
 \\
 &m' = - b
\end{align}
for $f_{s_n} = C_{s_n}' \delta (p - p_0)$.

The photon emission from Higgs is suppressed by an additional power of $(p_0- p)/p_0$.
This process is subdominant as can observed from Fig.~\ref{fig:flow1}.

\subsubsection{Case with $e_f$ production}

For $s_{n+1} = e_f$, we generally write
\begin{align}
  &- \tilde{\gamma}_{e_f \to B e_f}
  \qty[ \int_0^{1}  \dd x \frac{1 + (1-x)^2}{x^{1/2}(1-x)^{1/2}} \,
    f_{e_f}(p)
    -
  \int_0^{x_{\rm max}-1} \dd x \frac{(1+x)^2 + x^2}{(1+x)^{1/2}x^{1/2}} \,
    f_{e_f}(p(1+x))
    ]
    \nonumber\\
    &\qquad\qquad\qquad\qquad\qquad\qquad\qquad\qquad\qquad\qquad\qquad +
    \tilde{\gamma}_{B \to e_f} \int_0^{x_{\rm max}-1}  \frac{\dd x}{x^{b}} \,
    f_{B}(p(1+x))
    \approx 0,
\end{align}
where $\tilde{\gamma}_{B \to e_f}$ and $\tilde{\gamma}_{e_f \to B e_f}$ are given by \eqref{eq:tildegatoqs} and \eqref{eq:tildeetoBe}, respectively.
Here we used the fact that only a case with $s_n = B$ is present.
Taking
$f_{e_f} = C_{e_f} \qty( (p_0 - p) / p_0 )^{m'}$, we obtain
\begin{align}
 &C_{e_f} =
\frac{8}{11\pi}
\frac{\Gamma(1-b) \Gamma (1+m)}{\Gamma(2-b+m)}  \frac{\tilde{\gamma}_{B \to e_f}}{\tilde{\gamma}_{e_f \to B e_f}} C_B
 \\
 &m' = m + 1 - b
\end{align}
for $f_{B} = C_{B} \qty( (p_0 - p) / p_0 )^m$
or
\begin{align}
 &C_{e_f} =
\frac{8}{11\pi}
\frac{\tilde{\gamma}_{B \to e_f}}{\tilde{\gamma}_{e_f \to B e_f}} C_{B}'
 \\
 &m' = - b
\end{align}
for $f_B = C_{B}' \delta (p - p_0)$. Note that $b=1/2$ in this case.

%%%%%%%%%%%%%
\small
\bibliographystyle{utphys}
\bibliography{draft}
%%%%%%%%%%%%%

%%%%%%%%%%%%%
\end{document}